\documentclass[pra,twocolumn,aps,showpacs,grouped address]{revtex4-1} 

\usepackage{amsmath,amssymb,amsfonts}
\usepackage{graphicx}
\usepackage{siunitx}

\usepackage{hyperref}
\hypersetup{
    unicode=true,
    colorlinks=true,
    linkcolor=blue,
    citecolor=blue,
    urlcolor=blue
  }

\begin{document}
\title{A versatile dual-species Zeeman slower for caesium and ytterbium}
\author{S.A.Hopkins}
\email{s.a.hopkins@durham.ac.uk}
 \affiliation{Joint Quantum Centre (JQC) Durham-Newcastle, Department of Physics, Durham University, South Road, Durham, DH1 3LE, United Kingdom.}
\author{K. Butler}
\affiliation{Joint Quantum Centre (JQC) Durham-Newcastle, Department of Physics, Durham University, South Road, Durham, DH1 3LE, United Kingdom.}
\author{R. Freytag}
\affiliation{Centre for Cold Matter, Blackett Laboratory, Imperial College London, Prince Consort
Road, London, SW7 2AZ, United Kingdom.}
\author{A. Guttridge}
\affiliation{Joint Quantum Centre (JQC) Durham-Newcastle, Department of Physics, Durham University, South Road, Durham, DH1 3LE, United Kingdom.}
\author{S. Kemp}
\affiliation{Joint Quantum Centre (JQC) Durham-Newcastle, Department of Physics, Durham University, South Road, Durham, DH1 3LE, United Kingdom.}
\author{E.A. Hinds}
\affiliation{Centre for Cold Matter, Blackett Laboratory, Imperial College London, Prince Consort
Road, London, SW7 2AZ, United Kingdom.}
\author{M.R. Tarbutt}
\affiliation{Centre for Cold Matter, Blackett Laboratory, Imperial College London, Prince Consort
Road, London, SW7 2AZ, United Kingdom.}
\author{S.L. Cornish}
\affiliation{Joint Quantum Centre (JQC) Durham-Newcastle, Department of Physics, Durham University, South Road, Durham, DH1 3LE, United Kingdom.}

\date{\today}

\begin{abstract}
We describe the design, construction  and operation of a versatile dual-species Zeeman slower for both Cs and Yb, which is easily adaptable for use with other alkali metals and alkaline earths. With the aid of analytic models and numerical simulation of decelerator action, we highlight several real-world problems affecting the performance of a  slower and discuss effective solutions. To capture Yb into a magneto-optical trap (MOT), we use the broad $^1S_0$ to $^1P_1$ transition at 399 nm for the slower and the narrow $^1S_0$ to $^3P_1$ intercombination line at 556 nm for the MOT. The Cs MOT and slower both use the D2 line ($6^2S_{1/2}$ to $6^2P_{3/2}$) at 852 nm. We demonstrate that within a few seconds the Zeeman slower loads more than $10^9$ Yb atoms and $10^8$ Cs atoms into their respective MOTs. These are ideal starting numbers for further experiments on ultracold mixtures and molecules. 

\end{abstract}


\maketitle

\section{Introduction}\label{sec:Intro}

The production of ultracold, heteronuclear, diatomic molecules in their ground state by coherently combining two different laser-cooled atomic species is currently an active research area \cite{Ni2008,Takekoshi2014,Molony2014,Park2015} because of the potential for quantum information processing \cite{DeMille2002,Carr2009,Krems2009CM,Ortner2011,BrickmanSoderberg2009,Baranov2012,Wall2013}, for cold chemistry \cite{DeMiranda2011,Krems2008,Ospelkaus2010}, the exploration of strongly-interacting quantum systems \cite{Wall2009,Micheli2007,CapogrossoSansone2010,Buechler2007} and precision measurement \cite{Hudson2011,ACME2014,Flambaum2007,Isaev2010}. Heteronuclear molecules manifest an electric dipole moment when a direction is imposed by an electric field \cite{Herzberg1950}, allowing study of long-range anisotropic interactions. If such polar molecules are loaded into an optical lattice, quantum simulation of  lattice-spin models associated with many unsolved problems becomes possible \cite{DeMille2002,Carr2009,Krems2009CM,Ortner2011,BrickmanSoderberg2009,Baranov2012}. The range of Hamiltonians and models of interest may be further extended \cite{Georgescu2014,Micheli2006,Lewenstein2007} by creating molecules that also have a magnetic dipole moment such as a diatomic pairing of an alkali metal with an alkaline earth, where the magnetic moment is associated with the remaining unpaired outer electron. A small number of groups, including ourselves, have constructed experiments with such studies in mind \cite{Hara2013,Khramov2014,Borkowski2013,Pasquiou2013}.

Owing to the novelty and exacting nature of such experiments, combined with the difficulty of calculating ab initio the molecular properties, it is not yet clear which specific molecules will be most conducive to the above stated goals. However the variety of diatomic molecules that are potentially realisable using laser-cooling methods will each offer unique properties for experimental study \cite{Wall2013}. Accordingly we have chosen to investigate the previously untried mixture of Cs and Yb for three main reasons: Firstly, routes to quantum degeneracy have been established for both species individually \cite{Weber2003a,Takasu2009}. Secondly, natural Yb consists of seven isotopes including five bosons and two fermions, thus allowing production of either bosonic or fermionic molecules with caesium and expanding the opportunities for finding a molecule with favourable properties. Thirdly, a novel Feshbach resonance mechanism has been predicted \cite{Brue2013} for ultracold collisions between Cs and Yb and it has been shown that the scaling of Cs-Yb collisional properties with isotopic mass should lead to this occurring for at least one isotope at an experimentally achievable magnetic field \cite{Zuchowski2010,Brue2013}. Such resonances may be useful as part of the route to combine cold Cs and Yb into molecules by magneto-association.

A first step for our dual-species experiment is to load sufficient numbers of atoms of both species into overlapped or adjacent magneto-optical traps (MOTs) in a region of ultra-high vacuum where evaporative cooling to degeneracy may take place.  We chose to accomplish this using Zeeman slowing \cite{Metcalf1982}, a well-established technique that both decelerates and cools atoms effusing from an atomic source using laser radiation resonant with a strong atomic transition. The Doppler shift of the laser frequency, which changes as the atom decelerates, is compensated by the Zeeman shift induced by a tailored magnetic field. Zeeman slowing of a single species has been an important tool \cite{Bagnato1989,Barrett1991,Molenaar1997,Dedman2004,Ovchinnikov2007,Dammalapati2009,Marti2010,Hill2012,ParisMandoki2014,Zhao2014,Bowden2015} for the field of ultracold atom trapping, typically enabling the collection in a few seconds of around $10^9$ atoms in a MOT.

When two species are required, it is convenient to slow them in the same `dual-species' Zeeman decelerator \cite{Okano2010,Marti2010,ParisMandoki2014,Pasquiou2013,Bowden2015} as this both saves space and avoids duplication of equipment. However, if the two species have significantly different properties as is the case for Cs and Yb (see Table \ref{tab:ElementProperties}) this requires careful design.  For a Yb MOT, the sensible choice for the cooling transition is a weak transition at 556\,nm which offers a low Doppler temperature of $\SI{4.4}{\micro\kelvin}$. However, this transition brings an associated problem: the MOT has a capture velocity of only a few metres per second, which is challenging to achieve with a slower. An optimally designed slower will capture a greater fraction of the flux from an oven with less laser power, resulting in faster loading of greater numbers of atoms, longer oven lifetimes and less contamination of the science chamber by unused flux.

In this paper, we gather together all the theory, data and design criteria relevant for building a versatile Zeeman slower, designed for Yb and Cs but capable of slowing many common laser-cooled species such as the alkali metals, alkaline-earths and other divalent atoms. With the aid of analytic models and numerical simulation of decelerator action, we highlight several real-world problems of slowing and discuss their solutions. Finally we demonstrate that within a few seconds the Zeeman slower loads MOTs of more than $10^9$ Yb atoms and $10^8$ Cs atoms, yet requires only 60\,mW of laser power for Yb and 3\,mW for Cs. 

In section \ref{sec:BasicModel} we review the simplest analytic model of Zeeman slowing and then set out the immediate design implications of this model for the alkali metals Li, Na, K, Rb, Cs and two divalent atoms Sr and Yb. In section \ref{sec:CsYb} we focus in more detail on requirements specific to our Cs and Yb slower.
In \ref{sec:RealModel} we review a more complete analytic  model \cite{Bagnato1989,Napolitano1990,Molenaar1997,Ovchinnikov2007} which provides further insight into Zeeman slowing.  In addition, we extend the theory of Zeeman slowing, reserving the details to Appendix A.
We then describe our numerical simulation which incorporates various real-world effects that are unavoidable in practice but not included in the analytic models. 
In \ref{sec:SimulationResults} we present the results of the simulation, leading to an optimised final design, and in \ref{sec:Fabrication} we provide practical details of our apparatus such as coil geometries, fabrication techniques and laser systems. Finally in section \ref{sec:Results} we present experimental data showing the successful use of our dual-species slower for caesium and ytterbium before concluding.

\section{ Basic model of Zeeman slowing}\label{sec:BasicModel}
\begin{figure}
\includegraphics[width=\columnwidth]{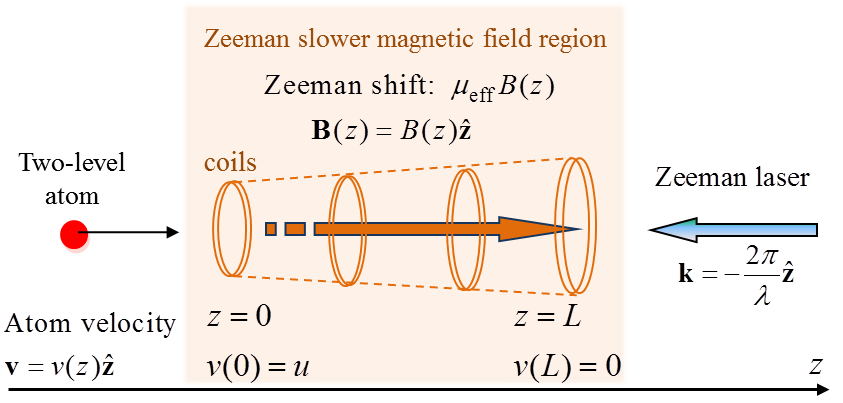}
\caption{A basic model of Zeeman slowing: A two-level atom of mass $m$ enters the Zeeman slower magnetic field region with initial velocity $\mathbf{v}(z) = u\mathbf{\hat{z}}$ along the $+z$-axis. The atomic transition has angular frequency $\omega_{\text{0}}$ and spontaneous decay rate $\Gamma$. The atom is decelerated by a counter-propagating plane wave with angular frequency $\omega_{\text{L}}$, wavelength $\lambda$, wavevector $\mathbf{k} = -k\mathbf{\hat{z}}$ and intensity $s$ in units of the saturation intensity, i.e. $s=I/I_{\text{sat}}$. In the frame of the atom the laser frequency is Doppler-shifted to  $\omega_{\text{L}} + kv(z)$.  A spatially varying magnetic field, $\mathbf{B}(z) = B(z)\mathbf{\hat{z}}$, acts in the Zeeman slower region from $z = 0$ to $z = L$, producing a Zeeman shift of the atomic transition frequency to  $\omega _{0} + \mu_{\text{eff}} B(z)/\hbar$.  Here we have defined an effective magnetic moment $\mu_{\text{eff}} = (m_eg_e - m_gg_g)\mu_{\text{B}}$ where $m_e$, $m_g$ are the magnetic quantum numbers, $ g_e$, $g_g$ are the Land$\acute{e}$ g-factors of the ground and excited states and $\mu_{\text{B}}$ is the Bohr magneton. Note that $\mu_{\text{eff}}$ may be positive or negative depending on the chosen two-level transition. The magnetic field is tailored so that the Zeeman shift just compensates the changing Doppler shift and hence the atom's speed can be efficiently reduced from $v(0) = u$ to $v(L) = 0$.}
\label{fig:Notation}
\end{figure}

We start by reviewing a basic model of a Zeeman slower as often presented in text books \cite{Foot2004}. Figure\,\ref{fig:Notation} and its caption define our notation. In the frame of the atom, the effective detuning $\delta (z,v)$ of the atomic transition is the difference between the Doppler shifted laser frequency $\omega_{\text{L}} + kv(z)$ and the Zeeman-shifted transition frequency $\omega _{0} + \mu_{\text{eff}} B(z)/\hbar$ so 
\begin{equation}
\delta(z,v) = \Delta+kv(z)-\mu_{\text{eff}} B(z)/\hbar,
\label{eq:LocalDetuning}
\end{equation}
where $\Delta = \omega_{L} - \omega_{0}$ is the fixed laser detuning in the lab frame. The resonance condition for an idealised slower is then simply $\delta(z,v) = 0$, which leads to the required magnetic field profile in terms of the speed of the atom:
\begin{equation} 
B(z) = \frac{\hbar}{\mu_{\text{eff}}}\Big(\Delta + kv(z)\Big).
 \label{eq:ResonanceCondition}
\end{equation}
  
\begin{figure}
\includegraphics[width=\columnwidth]{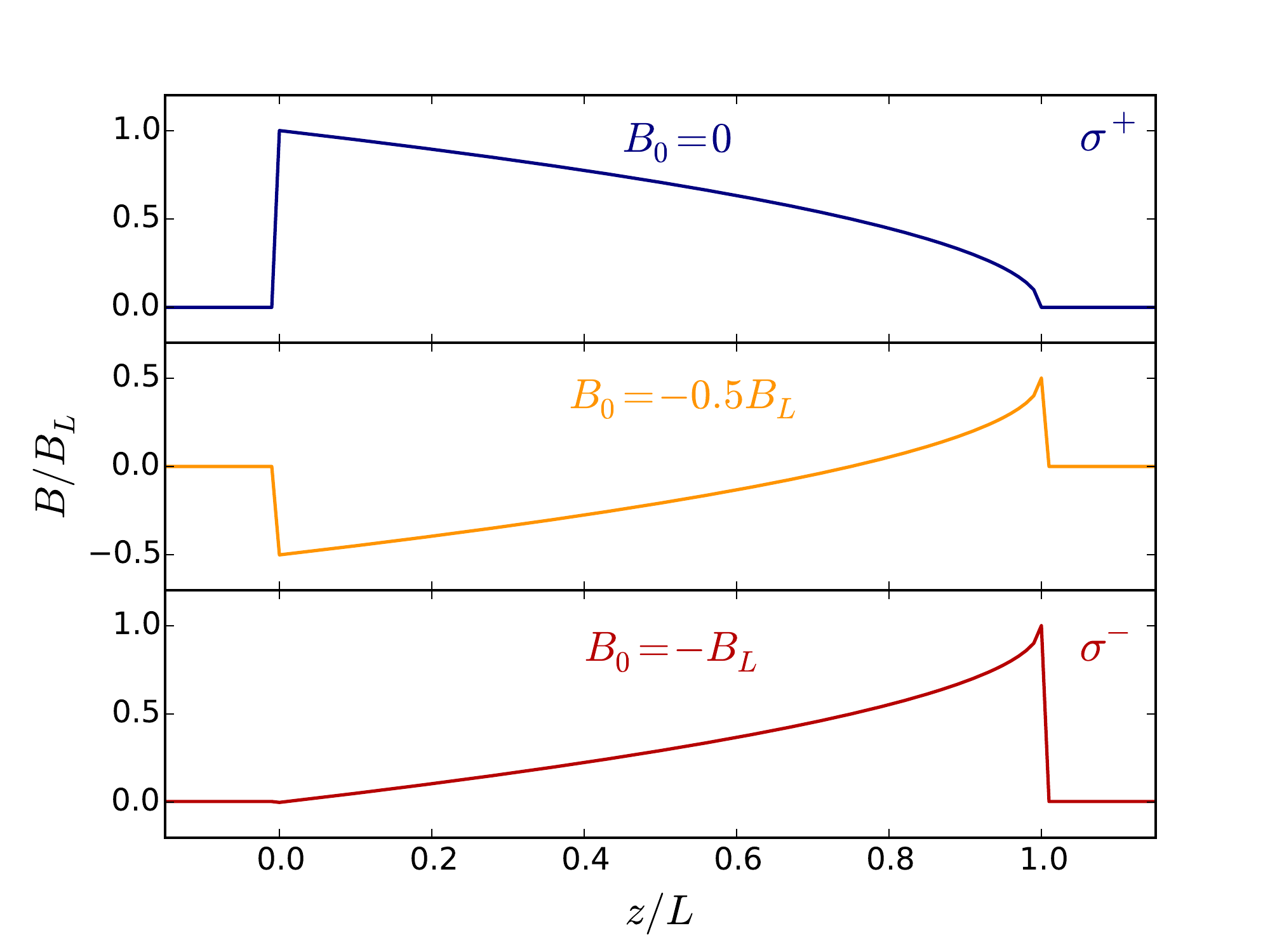}
\caption{Examples of possible magnetic field profiles $B(z)$, in units of $|B_L|$ vertically and $L$ horizontally. In the top row, $\mu_{\text{eff}}$ is positive and such slowers, with a decreasing-field profile, are generally referred to as $\sigma^{+}$ slowers because the laser drives $\sigma^{+}$ transitions. Conversely in the bottom row, $\mu_{\text{eff}}$ is negative and hence these increasing-field slowers are usually referred to as $\sigma^{-}$ slowers. The three rows correspond to: top $B_0=0$, middle $B_0 = -0.5B_L$ and bottom $B_0=-B_L$ or equivalently $\Delta=0, \Delta=-0.5ku, \Delta=-ku$. The field can change sign at some point along the slower (middle row); we refer to such designs as `zero-crossing' slowers.}
\label{fig:AlphaConfigs}
\end{figure}

Standard laser-cooling theory for a two-level atom gives the deceleration $dv/dt$ due to the laser beam as
\begin{equation}
 \frac{dv}{dt} = -\frac{1}{m}\frac{s}{1 + s + 4\delta^2/\Gamma^2}\frac{\hbar k\Gamma}{2}. 
 \label{eq:ScatteringRate}
 \end{equation}
Here $s=I/I_{\text{sat}}$ is the laser intensity in units of the saturation intensity of the transition, treated as a constant (a plane wave) in this basic model. Assuming we can create the field profile, $B(z)$, of Eq. (\ref{eq:ResonanceCondition}) exactly so as to maintain $\delta(z,v) = 0$, Eq. (\ref{eq:ScatteringRate}) simplifies to a constant deceleration
\begin{equation}
 \frac{dv}{dt} =- \eta \frac{\hbar k\Gamma}{2m} =- \eta a_{\text{max}},
 \label{eq:Acceleration}
 \end{equation}
where $a_{\text{max}} = \hbar k\Gamma/2m$ is the maximum possible magnitude of acceleration for a fully saturated transition and the `deceleration parameter' $\eta = s/(1+s)$ is the fraction of $a_{\text{max}}$ actually deployed. 
The velocity profile then follows from Eq. (\ref{eq:Acceleration}): $v(z) = \sqrt{u^2-2\eta a_{\text{max}}z}$ where $u$ is the initial speed. The atoms will be slowed to a stop after a distance $L$ given by
\begin{equation}
L=\frac{u^2}{2\eta a_{\text{max}}},
 \label{eq:Length}
 \end{equation}
and we may write
\begin{equation}
v(z) = u\sqrt{1-z/L}.
 \label{eq:vprofile}
 \end{equation}
Substitution of Eq. (\ref{eq:vprofile}) into (\ref{eq:ResonanceCondition}) then gives the required magnetic field profile as a function of $z$:
\begin{equation}
B(z) = B_L\sqrt{1-z/L} + B_0, 
 \label{eq:Bprofile}
 \end{equation}
where we have made the substitutions $B_L = \hbar k u/\mu_{\text{eff}}$ and $B_0 = \hbar\Delta/\mu_{\text{eff}}$, which have signs dependent on $\mu_{\text{eff}}$ and $\Delta$.
The first term on the right hand side of Eq. (\ref{eq:Bprofile}) contains the spatial dependence of $B(z)$, i.e. a total change of magnitude $|B_L|$. The second term $B_0$ is a constant offset field proportional to the laser detuning which can in principle take any value. It is also simply the field at the slower exit where $z=L$. The freedom to choose the sign of $\mu_{\text{eff}}$ and the magnitude of $\Delta$ enables different field configurations for Zeeman slowers and we have illustrated some generic field profiles in Fig.\,\ref{fig:AlphaConfigs}. 

\begin{figure}
\includegraphics[width=\columnwidth]{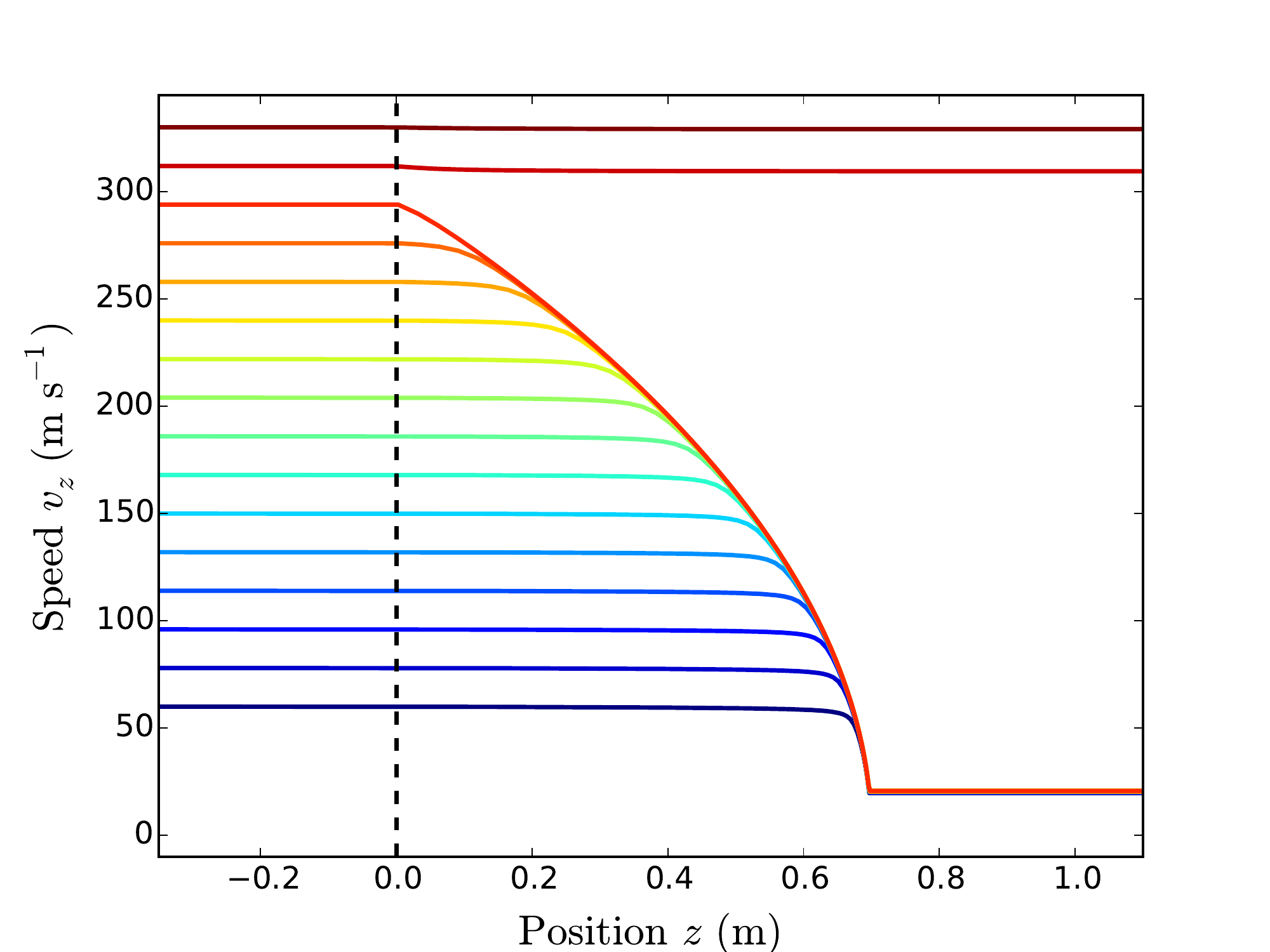}
\caption{Atom trajectories in a Yb slower designed with capture speed $u=300$\:m\,s$^{-1}$, simulated in accordance with the basic model. The slowing action is artificially confined to the region $0 \leq z \leq z_{\text{rel}}$ using the magnetic field defined in Eq. (\ref{eq:Bprofile}) and shown in any of the profiles of Fig.\,\ref{fig:AlphaConfigs}. Sixteen atoms are injected along the $z$-axis with a uniform spread of longitudinal speeds and we see that atoms entering with speeds above 300\:m\,s$^{-1}$ escape, but all other atoms converge onto the capture envelope until finally released at the chosen release speed, in this example $v_{\text{rel}}=20\:\text{m\,s}^{-1}$.} 
\label{fig:SimulatorIdeal}
\end{figure}

In Fig.\,\ref{fig:SimulatorIdeal} we show a set of velocity profiles, simulated in accordance with the basic model presented so far. Any atoms entering with speeds $u'<u$ are initially out of resonance with the laser, so will progress forward at near-constant speed and may then be captured by the section of the field profile from $z=L(1-(u'/u)^2)$ to $z=L$; they converge onto the velocity profile of Eq. \ref{eq:vprofile}. Atoms entering with speeds greater than $u$ do not come into resonance at any stage and are not significantly slowed. Hence from now on we refer to $u$ as the `capture speed' of the slower and the trajectory of an atom entering with speed $u$ as the `capture envelope'.

Slowers are designed to release atoms at a speed less than the MOT capture speed, but large enough for them to progress successfully to the MOT capture region which is usually displaced from the end of the slower for technical reasons such as the positioning of coils and vacuum apparatus. In principle any desired exit speed $v_{\text{rel}}$ can be achieved by truncating the field profile at release position $z_{\text{rel}} = L(1-(v_{\text{rel}}/u)^2)$. In reality the slowing action cannot be switched off so abruptly, as will be discussed later. 

\subsection{Practical implications of basic model}
\label{subsec:FirstDesign}
Table \ref{tab:ElementProperties} lists the relevant physical parameters and derived quantities for seven common laser-cooled alkali metals and alkaline earths: Li, Na, K, Rb, Cs, Sr and Yb. In order to load large MOTs of about $10^9$ atoms in a few seconds, an effusive atomic oven with a vapour pressure around $10^{-3}$ mbar is required to produce a sufficiently high atomic flux into the entrance aperture of the slower \cite{Giordmaine1960,Ross1995}. The necessary oven temperatures (line (v) in Table \ref{tab:ElementProperties}) vary in the range 100 -- 500\,$^\circ$C and the associated mean speeds (vi) are in the range 250 -- 700\;m\,s$^{-1}$ except for the light element Li. 
\begin{table*}[ht]
	\centering
		\begin{tabular}{llccccccccc}\hline\hline
					&Quantity 	     											& Symbol							& Units					& $^{6}$Li			& $^{23}$Na 			& $^{39}$K  	 & $^{87}$Rb    	& $^{133}$Cs		& $^{88}$Sr			& $^{174}$Yb		\\ \hline
		i			&Wavelength 													& $\lambda$						& nm						& 671						& 589							& 767				 	 & 780			  		& 852						& 461						& 399						\\ 
		ii		&Decay rate 													& $\Gamma/2\pi$				& MHz						& 5.9						& 9.8							& 6.0				 		 & 5.9			  	& 5.2						& 30.2					& 28.0						\\ 
		iii		&Maximum deceleration									& $a_{\text{max}}$		& km\,s$^{-2}$		& 1820					& 902							& 251					 & 108			  		& 57						& 984						& 548						\\ 
		iv		&Saturation intensity									& $I_{\text{sat}}$		& mW\,cm$^{-2}$		& 2.5						& 6.3							& 1.8				 	 & 1.6		  			& 1.1						& 43						& 63						\\
		v			&Oven temperature 										& $T_{\text{oven}}$		& $^\circ$C			& 440						& 230							& 155		 			 & 120			  		& 105						& 460						& 400						\\ 
		vi		&Mean speed 													& $u_{\text{mean}}$		& m\,s$^{-1}$			& 1580					& 675							& 481				 	 & 306			  		& 245						& 419						& 285						\\ 
		vii		&Minimum length												& $L_{\text{min}}$		& m							& 1.03					& 0.38						& 0.69				 & 0.65			  		& 0.78					& 0.13					& 0.11					\\ 
		viii	&Magnetic field span									& $B_L$								& G					& 1690					& 819							& 448				 	 & 281			  		& 205						& 650						& 511						\\
		ix		&Scattering events										& $N_{\text{abs}}$		& $\times10^3$	& 16						& 23							& 36				 	 & 52				  		& 69						& 43						& 50						\\ 
		x			&Relative deceleration parameter			& $\eta_{\text{r}}$		& -							& 1							& 1.55						& 9.4					 & 22.7						& 51.3					& 0.87					& 1.17					\\	\hline\hline
		\end{tabular}
	\caption[Properties of alkali metals and alkaline earths]{Comparison of atomic properties \cite{NIST2003,CRC2003} and relevant derived quantitites for Zeeman slowing for alkali metals and two alkaline earths. (i-iv) The wavelength, decay rate, maximum deceleration and saturation intensity are those associated with the strongest available slowing transition for each species. In row (iv) $I_{\text{sat}}=\pi h c \Gamma/(3 \lambda^3)$. (v) The oven temperatures are those necessary to produce a vapour pressure of about $10^{-3}$\,mbar. (vi) The mean speed $u_{\text{mean}}=(8k_{\text{B}}T_{\text{oven}}/\pi m)^{1/2}$. (vii) The minimum length calculated from Eq. (\ref{eq:Length}) with $\eta=0.67$. (viii) The magnetic field span $|B_L| = |\hbar k u/\mu_{\text{eff}}|$. (ix) The number of scattering events is the initial atomic momentum divided by the photon momentum. (x) Relative deceleration parameters \cite{Wille2009} $\eta_{\text{r}}$ as described in the main text.}  
	\label{tab:ElementProperties}
\end{table*}

The minimum lengths (vii) are obtained from Eq. (\ref{eq:Length}): $L=u^2/(2\eta a_{\text{max}})$ where $a_{\text{max}}$ is fixed by the wavelength and decay rate of the slowing transition and we use $\eta=0.67$ as an appropriate maximum value. As the length scales as $u^2$ we compromise with a capture speed $u=u_{\text{mean}}$, where $u_{\text{mean}}$ is the mean speed of atoms effusing from the oven; this implies that the most of the lower half \cite{ZeemanMedian} of a Maxwellian speed distribution can be captured. We see that Li and Cs require the longest slowers and Sr and Yb the shortest; this is simply due to the particular combination of oven effusion speed and maximum deceleration for each element. A slower may be too short to reach the desired efficiency, here defined as $u/u_{\text{mean}}$, but one that is `too long' is easily compensated by a reduction of $\eta$ via the laser intensity. In row (viii) we give the associated magnetic field span $|B_L|$ and we see that it is generally of order a few hundred gauss, hence large coils with hundreds of amp-turns are usually constructed. Alternative schemes have been successfully demonstrated, e.g. deploying arrays of permanent magnets \cite{Cheiney2011,Hill2012,Lebedev2014,Parsagian2015} or a single winding with a variable pitch \cite{Bell2010}. For Li and Na the required field span becomes rather large which stems ultimately from their small mass and associated large oven effusion speed. Row (ix) gives the number of scattering events needed to bring the atom from $u_{\text{mean}}$ to zero and is typically a few $10^4$. The momentum diffusion from this many spontaneous emissions results in final transverse rms speeds around 0.5\,m\,s$^{-1}$.

The final row (x), gives relative deceleration parameters $\eta_{\text{r}}$ as originally described in Ref. \cite{Wille2009} and calculated as follows. We envisage a slower with some fixed length $L$ and field span $B_L$ and determine the necessary deceleration parameter $\eta_{\text{i}}$ for each element $i$ using Eqs. (\ref{eq:Acceleration}), (\ref{eq:Length}) and (\ref{eq:Bprofile}). All the $\eta_{\text{i}}$ have the same dependence on $L$ and field span $B_L$ so for any two elements $a$ and $b$ we can form the ratios $\eta_{\text{r}}=\eta_{\text{a}}/\eta_{\text{b}}=m_a \mu_{\text{eff}\_a}^2 \lambda_a^3 \Gamma_a^{-1} /m_b \mu_{\text{eff}\_b}^2 \lambda_b^3 \Gamma_b^{-1}$  which depend only on atomic properties. In row (x) we have taken element $b$ to be Li. The interpretation is that if any two $\eta_{\text{r}}$ are within a factor of say 0.5 - 2 of each other, e.g. Li and Yb, then both species can be slowed simultaneously and with good efficiencies using the same magnetic field and length. When two $\eta_{\text{r}}$ parameters are related by a larger factor, the field will either need to be switched for sequential loading of each species or alternatively one species slowed with a poor capture efficiency.

Two observations regarding Table \ref{tab:ElementProperties} are i) that a slower long enough to slow Li or Cs can be used to slow any of the listed species and ii) that the two elements Cs and Yb fall towards opposite extremes and our experiment therefore serves as a generic example for dual-species Zeeman slowers.

We now consider the counter-propagating laser (`Zeeman laser') of the slower. The deceleration parameter was defined as $\eta=s/(1+s)$ where $s=I/I_{\text{sat}}$. $I$ is the laser intensity and $I_{\text{sat}}$ is the saturation intensity. In practice $\eta$ is restricted to values less than 0.67 mainly because higher values reap little benefit whilst requiring a large increase in laser power. Furthermore it has been shown \cite{Molenaar1997} that the value $\eta=0.5$ gives the optimal damping of the relative speeds onto the design speed profile and hence the narrowest spread of speeds at the exit. Keeping $\eta\leq 0.67$ implies that the required laser intensity will not be more than $2I_{\text{sat}}$ which in turn implies, see Table \ref{tab:ElementProperties}, that only a few mW\,cm$^{-2}$ are needed for the alkali metals. By contrast, the alkaline earths need of order 100\,mW\,cm$^{-2}$. 

The laser detuning $\Delta$ and the offset field $B_0$ constitute a single free parameter as they are proportional to each other. The value actually used however is strongly influenced by experimental considerations. For example, we avoid a design where $B_0 = 0$ (as in the top row of Fig.\,\ref{fig:AlphaConfigs}) because this corresponds to $\Delta = 0$, meaning that the Zeeman laser is resonant with the atoms trapped in the MOT. Since the Zeeman laser passes through the MOT this would exert a strong scattering force on the weakly trapped Yb atoms and we could not change the Zeeman laser without strongly influencing the MOT \cite{Barrett1991,Molenaar1997,Dammalapati2009}. 
A design with a large detuning at the exit, as depicted in the bottom row of Fig.\,\ref{fig:AlphaConfigs}, can be advantageous \cite{Barrett1991}. This is because the pushing of the MOT is much reduced and the release speed of the slower is more easily controlled because there is a large change in Zeeman shift at the exit. However the large end field must be reduced to zero over the short distance to the MOT so as not to perturb the small MOT magnetic field gradient. Commonly the `zero-crossing' slower is used as a compromise.

For atoms with hyperfine energy level structure (all the alkali metals), a suitable repumping laser must be overlapped with the Zeeman laser to minimise losses from the cycling transition due to off-resonant optical pumping. In the case of potassium \cite{Wille2009}, the excited state hyperfine levels are so closely spaced that the scattering of repumping and cooling light both play a role in the slowing. A complication is that the effective magnetic moment $\mu_{\text{rep}}$ for the repumping transition is not usually equal to $\mu_{\text{eff}}$ for the slowing transition, and consequently the repumper detuning and repumping rate vary along the slower.  

\section{Practical considerations for C\lowercase{s} and Y\lowercase{b}}\label{sec:CsYb}

\begin{figure}
\includegraphics[width=\columnwidth]{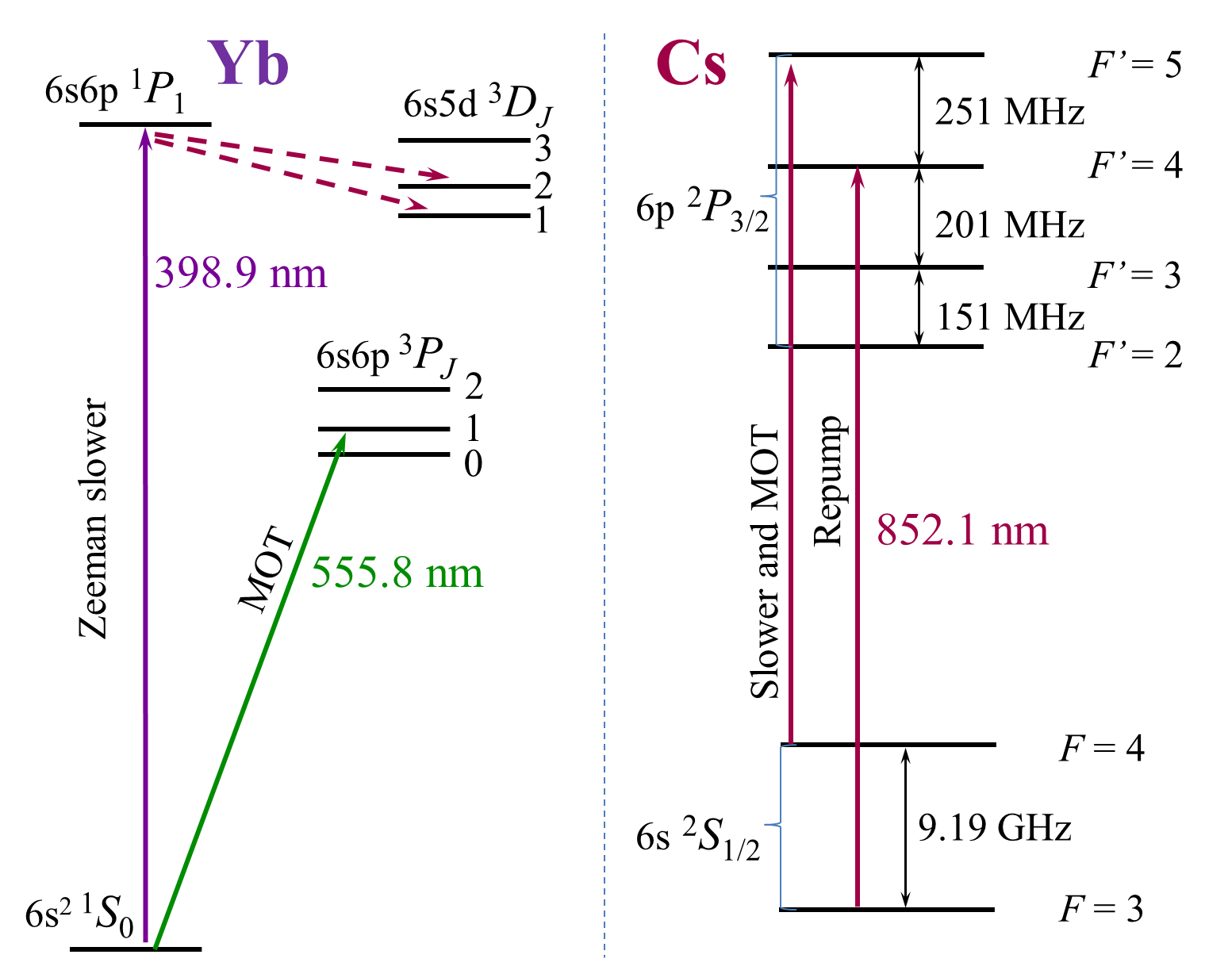}
\caption{Relevant energy levels of Yb (left) and Cs (right). In Yb there is a strong transition (violet) $^1S_0$ to $^1P_1$ at 398.9\,nm with $\Gamma/2\pi = 28$\,MHz and a weaker transition (green) $^1S_0$ to $^3P_1$ at 555.8\,nm with $\Gamma/2\pi = 182$\,kHz. The 399\,nm transition has a weak branching decay to the D levels. In Cs the cooling and repumping transitions (red) operate on the $D_2$ lines $6^2S_{1/2}$ to $6^2P_{3/2}$ at 852.1\,nm with hyperfine structure as shown.}
\label{fig:Yb&Cs}
\end{figure}

We now focus upon the specific requirements of our dual-species Cs-Yb experiment. Fig.\,\ref{fig:Yb&Cs} shows two transitions in Yb suitable for cooling and trapping: a strong transition at 399~nm and a weaker intercombination line at 556\,nm. The only practical choice for the Zeeman slower is 399~nm (a 556 nm slower would need to be 27 m long!). For the MOT, we utilise the 556 nm transition because it is closed, whereas the 399 nm transition excites a weak decay to the D levels that limits the maximum MOT number to around $10^6$ atoms \cite{Loftus2000,Honda1999}. Furthermore, the narrow 180 kHz width of the 556 nm line results in a very low Doppler temperature $T_{\text{D}}$ of $\SI{4.4}{\micro\kelvin}$ which facilitates the subsequent transfer to an optical trap. However this narrow linewidth also results in a small MOT capture velocity \cite{Kuwamoto1999}, which we find by numerical simulation of the MOT to be $\sim7$~m\,s$^{-1}$ for our highly power-broadened MOT beams (each has 13 mW in a $1/e^2$ diameter of 25 mm leading to $s=I/I_{\text{sat}}\approx 40$).

\begin{figure}
\includegraphics[width=\columnwidth]{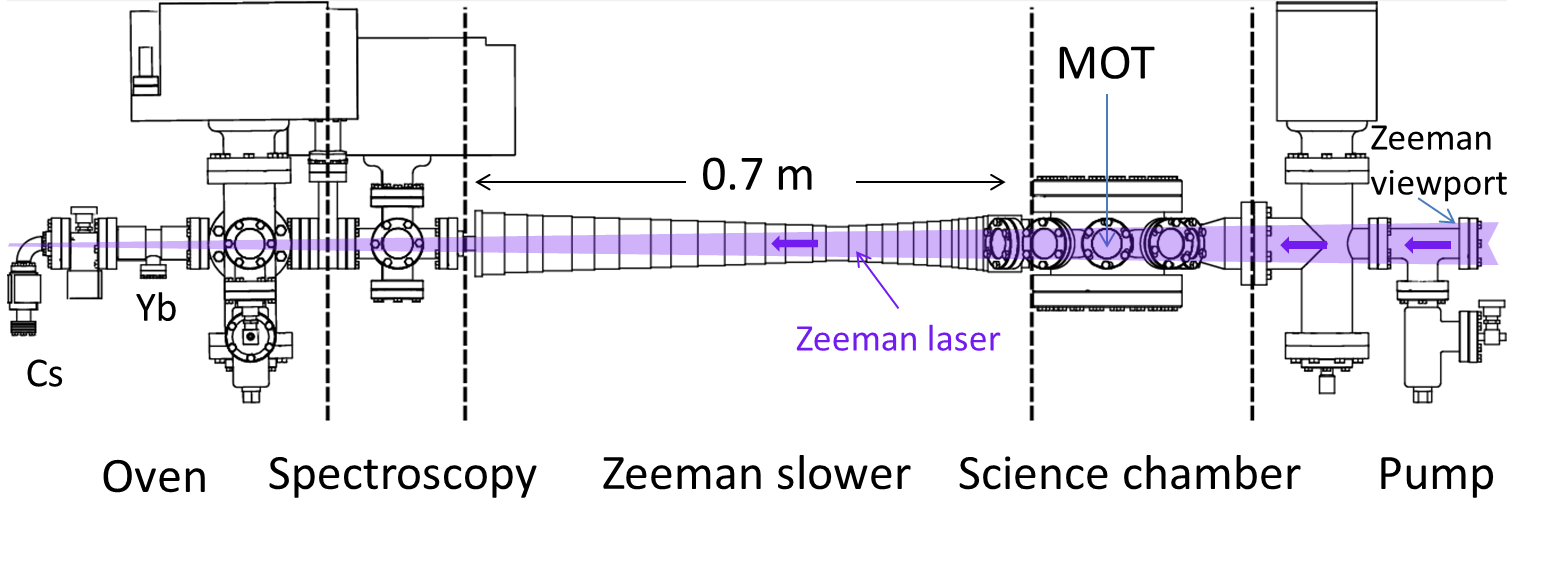}
\caption{Overview (to scale) showing the main sections of our  apparatus. The Zeeman laser beams, indicated in violet, enter with large diameters through the Zeeman viewport and focus towards a point about 0.2\,m to the left of the oven.}
\label{fig:apparatus}
\end{figure}

Atoms released from the slower with speeds less than 5\,m\,s$^{-1}$  can fall too far under gravity to enter the MOT capture region. Furthermore, the slowest atoms diverge strongly from the end of the decelerator because of the transverse velocity spread that results from transverse heating (see Fig.\,\ref{fig:SimulatorResults}(b)).  It is therefore vital to minimise the distance from the slower exit to the MOT; in our design it is 12.75\,cm, constrained by our science chamber vacuum housing. At this distance, we need the slower to deliver Yb atoms at speeds between 2 and 7\,m\,s$^{-1}$. An overview of our vacuum apparatus is given in Fig.\,\ref{fig:apparatus} and described in detail in Ref. \cite{Kemp2015}.

For both species we chose the zero-crossing magnetic field profile depicted in the middle plot of Fig.\,\ref{fig:AlphaConfigs}. This corresponds to having a large detuning at the slower exit, and the field reversal allows the use of lower coil currents to create the total field span $B_L$. The minimum length of the slower is dictated by the stopping distance of Cs (see Table \ref{tab:ElementProperties}); we decided on a length of 0.7\,m, offering a reasonable capture efficiency $u/u_{\text{mean}}$ of 0.81 for Cs with $\eta_{\text{Cs}}$ = 0.5.  This length is greater than needed for Yb alone and it then follows that we can achieve a similar capture efficiency of $\approx1$ for Yb with a small value of $\eta_{\text{Yb}}$ of 0.128; this confers the advantage that a value of $0.15\,I_{\text{sat}}$ or 9\,mW\,cm$^{-2}$ is sufficient for the Yb laser intensity.   

\begin{figure}
\includegraphics[width=\columnwidth]{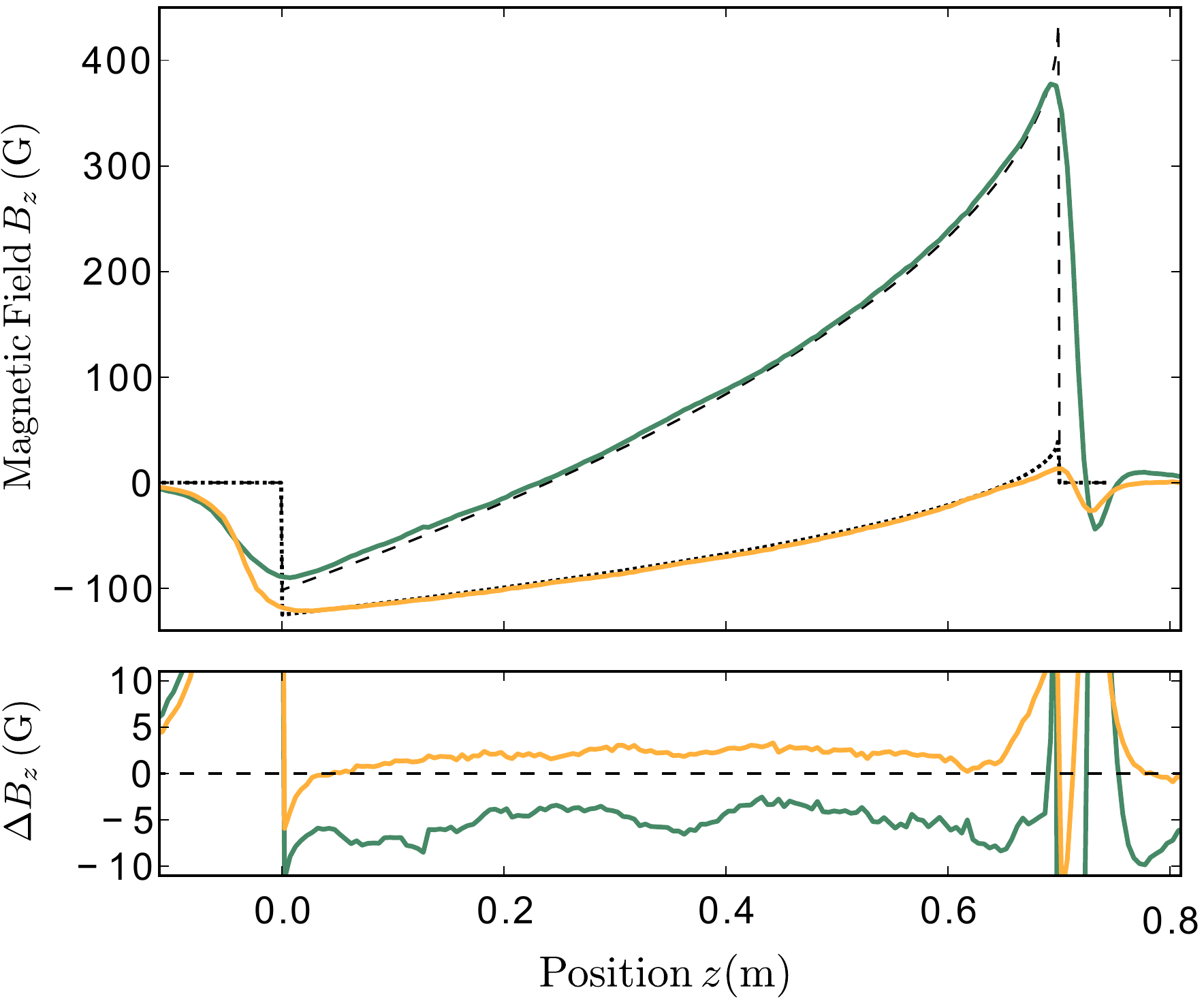}
\caption{Top: The measured axial magnetic fields of the Zeeman slower, plotted against position whilst running at Yb currents (green) and Cs currents (orange). The model fields of Eq. (\ref{eq:Bprofile}) for each element are shown by the dashed and dotted-dashed black lines for Yb and Cs respectively. The centre of the MOT is at 0.83 m. Bottom: The residuals between the measured and model fields.}
\label{fig:FieldProfiles}
\end{figure}

In Fig.\,\ref{fig:FieldProfiles} we show the differing magnetic field profiles required for Cs and Yb. To realise a dual-species slower, we designed a set of five coils which can produce both field profiles by switching the coil currents with the aim of loading Cs and Yb sequentially.

In order to prevent pushing of the delicate 556\,nm MOT by the strong 399\,nm Zeeman laser, this Yb scheme requires a large Zeeman laser detuning of at least 600\,MHz and a corresponding end field $B_0$ of 435\,G; the full field span $B_L$ is 537\,G. because of the narrow linewidth, the MOT magnetic field gradient is small, typically around 3\,G\,cm$^{-1}$ axially \cite{Kuwamoto1999}, so any residual Zeeman field must be well nulled over a short distance around 10\,cm. The main parameters of our chosen design are summarised in Table \ref{tab:ParameterSummary}.

We turn now to Cs. Our slower uses an end field of 42\,G with a field span of 167\,G. The MOT and slower both use light on the standard, circularly-polarised $F=4$ to $F'=5$ cycling transition, as indicated in Fig.\,\ref{fig:Yb&Cs}. The MOT capture velocity is about 40\,m\,s$^{-1}$, much less challenging for the slower than the 7\,m\,s$^{-1}$ required for Yb. In fact we can load Cs atoms directly from the oven even if the Zeeman fields are off, although only to around 1\% of the full load. One complication for Cs is its multi-level structure. We choose a Zeeman laser detuning around 50\,MHz below the cycling transition. A larger detuning causes too much atom loss via excitation of the $F'=4$ state even though a repumper co-propagates with the Zeeman laser. For the repumper, we find empirically that 50\% of the slowing power is sufficient with a soft optimum in the repumper detuning at 40\,MHz to the red of $F=3$ to $F'=4$.

\begin{table}[ht]
	\centering
		\begin{tabular}{ccccc}\hline\hline
			Quantity 	     						& Symbol												& Units					& Cs				& Yb 					\\ \hline
			Length 	     							& $L$														& m						& 0.7				& 0.7	 				\\ 
			Deceleration parameter		& $\eta$												& -							& 0.5				& 0.128				\\ 
			Capture speed 						& $u$														& m\,s$^{-1}$			& 200				&	300						\\ 
			Magnetic field span 			& $B_L$													& G					& 167				&	537						\\ 
			Magnetic field offset  		& $B_0$													& G					&	42				& 435					\\ 
			Laser detuning						& $\Delta$											& MHz						&	-59				& -609					\\ 
			Laser intensity						& $s$											& $I_{\text{sat}}$		&	1				& 0.147					\\ 
			Desired release speed			& $v_{\text{rel}}$							& m\,s$^{-1}$			&	15				& 5					\\ 
			Capture ratio							& $u/u_{\text{mean}}$						& -							&	0.81			& 0.98				\\  \hline
		\end{tabular}
	\caption{Design parameters stemming from the simple model as chosen for our Cs-Yb slower. In this model the laser is treated as a plane wave and its intensity $s$, in units of the relevant $I_{\text{sat}}$, is related directly to the deceleration parameter via $\eta=s/(1+s)$.}  
	\label{tab:ParameterSummary}
\end{table}

\section{Realistic Zeeman slowing of C\lowercase{s} and Y\lowercase{b}}\label{sec:RealModel}
To progress from the basic design elements so far presented towards a final design, we made use of two tools, described in this section. The first is a full analytic model \cite{Bagnato1989,Napolitano1990,Molenaar1997,Ovchinnikov2007} of Zeeman slowing and the second is a numerical simulation of the slower which incorporates various real-world physical effects not easily included in the analytic models. The analytic model helps to illuminate the results of the numerical simulation. We only summarise it here; a fuller account, including a modest extension, is provided in Appendix A.

The basic model developed in section \ref{sec:BasicModel} has the interesting shortcoming that its trajectories for atoms entering with $u'< u$ are all stable, while those with $u'\geq u$ are completely unstable. Further insight \cite{Dedman2004,Marti2010} may be found by rewriting $\eta$ in the form 
\begin{equation}
\eta \equiv \frac{s}{1+s} = \frac{\mu_{\text{eff}}}{\hbar k a_{\text{max}}}v(z)\frac{dB(z)}{dz},
 \label{eq:IdealBalance}
 \end{equation}
which follows straightforwardly from equations (\ref{eq:ResonanceCondition}) and (\ref{eq:Acceleration}), togther with $dv/dt=v dv/dz$.
This shows that, along the critical capture trajectory of the basic model (see Figue \ref{fig:SimulatorIdeal}), there is always an \emph{exact} balance between the laser intensity $s$, the velocity $v(z)$ and the field gradient $dB/dz$. In such a slower, if any of these three quantities fluctuates in the `wrong' direction from its local design value, (e.g. the laser intensity decreases, the velocity increases because of Poissonian noise in the scattering rate, or the field gradient increases because of local ripples), then the local deceleration becomes insufficient and the atom immediately escapes over the critical capture envelope and is lost. This problem is solved by increasing the laser intensity to $s'$, while keeping the field profile appropriate for $s$, as discussed in Appendix A. Then the atom follows a trajectory $v'(z)$, offset from the original capture trajectory by an amount $\epsilon = v(z) - v'(z)$, given by
\begin{equation}
 \epsilon\cong\frac{\Gamma}{2k} \bigg(\frac{s'-s}{s}\bigg)^{1/2}.
\label{eq:epsilonApprox}
\end{equation}
This is a good approximation except at the end of the slower, when $v$ becomes small. A more accurate (but less transparent) expression, previously unpublished, is also derived in Appendix A. The speed offset $\epsilon$ stabilises the new trajectory: if the atom speeds up it comes closer to resonance and is decelerated more; if the atom slows down it is decelerated less and catches up; notably the scattering rate per atom remains constant \cite{Bagnato1989}. This negative feedback mechanism makes Zeeman slowers robust by providing headroom against the various fluctuations described above. See Fig.\,\ref{fig:Stability} of Appendix A. 

\subsection{Numerical simulation of Zeeman slowing}\label{subsec:Simulation}

There are significant differences between a real slower and the models, some of which are potentially under our experimental control, such as the accuracy of the magnetic field profile or collimation of the atomic beam, whilst others are unavoidable such as the transverse heating. A numerical simulation of the slower allows us to study such effects. In the simulation the mean acceleration is 
\begin{equation}
 \bar{\textbf{a}}(\textbf{r},\textbf{v}) = \frac{1}{m}\frac{s(\textbf{r})}{1 + s(\textbf{r}) + 4\delta^2(\textbf{r},\textbf{v})/\Gamma^2}\frac{\hbar \textbf{k}(\textbf{r})\Gamma}{2}, 
 \label{eq:RealAcceleration}
 \end{equation}
where the position and velocity dependence of the terms indicate the inclusion of the Gaussian intensity profile of the Zeeman laser and the spread of radial positions and transverse velocity components of the atoms. The laser intensity profile is
\begin{equation}
I(\textbf{r}) = \frac{2P}{\pi w^2(z)}\text{exp}\bigg(\frac{-2\rho^2}{w^2(z)}\bigg).
 \label{eq:Gaussian}
 \end{equation}
Here $P$ is the power and ${\rho^2=x^2+y^2}$. The beam $1/e^2$ radius is ${w=w_0\sqrt{1+(z-z_0)^2/z_{\text{r}}^2}}$, where $z_{\text{r}}$ is the Rayleigh range, $w_0$ is the minimum spot size, and $z_0$ is the location of this minimum. The $k$-vector is ${\textbf{k}(\textbf{r}) = k(x/R,y/R,\sqrt{1-\rho^2/R^2})}$, where ${R=(z-z_0)+z_{\text{r}}^2/(z-z_0)}$ is the radius of curvature of the wavefronts.

The deceleration fluctuates about the mean value of Eq. (\ref{eq:RealAcceleration}) because the absorption rate is subject to Poissonian fluctuations and the associated spontaneous emissions also cause momentum diffusion. The realistic field profile $\textbf{B}(\textbf{r})$ which enters the detuning term is the one we have measured for our complete Zeeman slower: it varies smoothly with no discontinuous changes, it extends beyond the beginning and end of the coil windings and it contains small-scale ripples as a consequence of the discrete windings. Furthermore, because $\nabla\cdot\textbf{B}=0$ and $dB_z/dz\neq0$, the actual magnetic field has non-zero radial components \cite{Dammalapati2009,Muniz2015,ZeemanDivergence} along its length as well as at the coil ends. The relative impact of all these factors was studied. 

In the simulation, a small ensemble of atoms are injected into the slower, usually with an initial spread of radial positions and velocities approximating the output of our dual-species oven (see Fig.\,\ref{fig:SimulatorBadReal} caption). For each atom $i$ and time step $\Delta t$ the mean scattering rate at the location of the atom is calculated with due regard to the local Doppler and Zeeman shifts and then a random number is selected from a Poissonian distribution with that mean to give the actual scattering rate $R_i$. The number of scattered photons is $n_i=R_i\Delta t$ and the change in momentum due to absorptions is $n_i\hbar \textbf{k}$. An isotropic \cite{ZeemanDipolar} random walk with $n_i$ steps in 3D momentum space gives the momentum diffusion occurring during the time step. The position and velocity of each atom are propagated in 3D in each time step to build up a phase-space trajectory through the slower. Runs with 5000 steps each of 10\,$\si{\micro\second}$ duration converged with sufficient precision to solutions revealing the salient features.

We first tested the simulation by reproducing the basic model of section \ref{sec:BasicModel}, i.e. by using the constant deceleration of Eq. (\ref{eq:Acceleration}) without Poissonian fluctuations or momentum diffusion, with a uniform plane wave Zeeman laser, with atoms injected along the $z$-axis only and with zero transverse velocity component; the results of this test were shown earlier in Fig.\,\ref{fig:SimulatorIdeal}. 

\section{Results of numerical simulation} \label{sec:SimulationResults}

\begin{figure}
\includegraphics[width=\columnwidth]{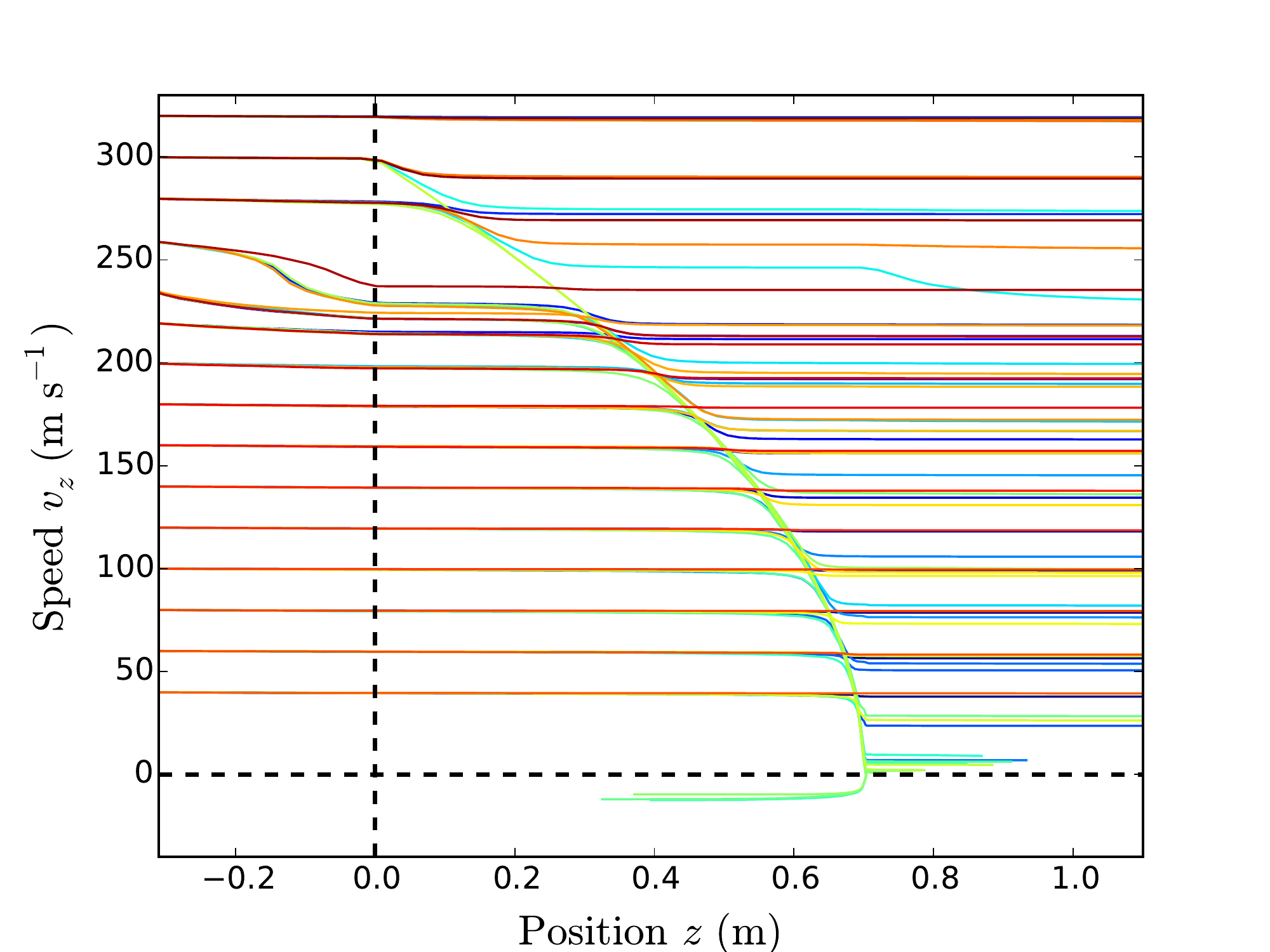}
\caption{A `bad' slower: Simulated trajectories ($v_z$ versus $z$) for atoms travelling through a realistic, non-optimal slower. Only a small percentage of the original atoms exit in the required velocity range. This simulation includes Poissonian fluctuations in the absorption and transverse momentum diffusion. We inject eighty atoms with a range of longitudinal speeds up to 340\,m\,s$^{-1}$ effusing from a 5\,mm diameter nozzle, 35\,cm in front of the slower entrance. The atoms are grouped into sixteen longitudinal entry speeds each containing five random  directions within a $\theta_{1/2}=10$ mrad entry cone. The Zeeman laser beam is a collimated Gaussian beam with a $1/e^2$ waist $\text{w}_0 = 7.5$\,mm. The magnetic field has a realistic profile identical to the green line of Fig.\,\ref{fig:FieldProfiles} except we have added ripples at the $\pm2$\,G level.  The curvature of trajectories for atoms entering with speeds around 240\,m\,s$^{-1}$ is due to the Doppler shift of those atoms bringing them temporarily into resonance with the Zeeman laser in the region of zero magnetic field before the slower entrance.}
\label{fig:SimulatorBadReal}
\end{figure}

Figure\,\ref{fig:SimulatorBadReal} provides a telling example of why simulation was valuable. This shows the results for a realistic Yb slower, with plausible parameters as far as the basic model goes, but which has far from optimal performance. Out of 80 atoms injected with speeds less than the capture speed, only 5\% end up in the desired exit velocity range from 2 -- 7\,m\,s$^{-1}$. The bulk of the atoms either escape too early from the capture envelope or are slowed to zero but then turn back towards the oven.

We now look more closely at the reasons for these losses and consider how to prevent them. The simulation revealed that the early escapes of  Fig.\,\ref{fig:SimulatorBadReal} are predominantly due to radially spreading atomic trajectories that enter regions of lower radial laser intensity $s'(\textbf{r})$. This reduces the available headroom $\epsilon$, see Eq. (\ref{eq:epsilonApprox}), and the trajectory becomes critically unstable. We used the simulation to compare possible solutions: the available headroom can be increased by a more uniform laser profile and/or more laser intensity; the spread of radial positions can be reduced by tighter collimation of the atomic beam and/or by application of transverse cooling light.  

\begin{figure*}
	\centering
	\includegraphics[width=1\linewidth]{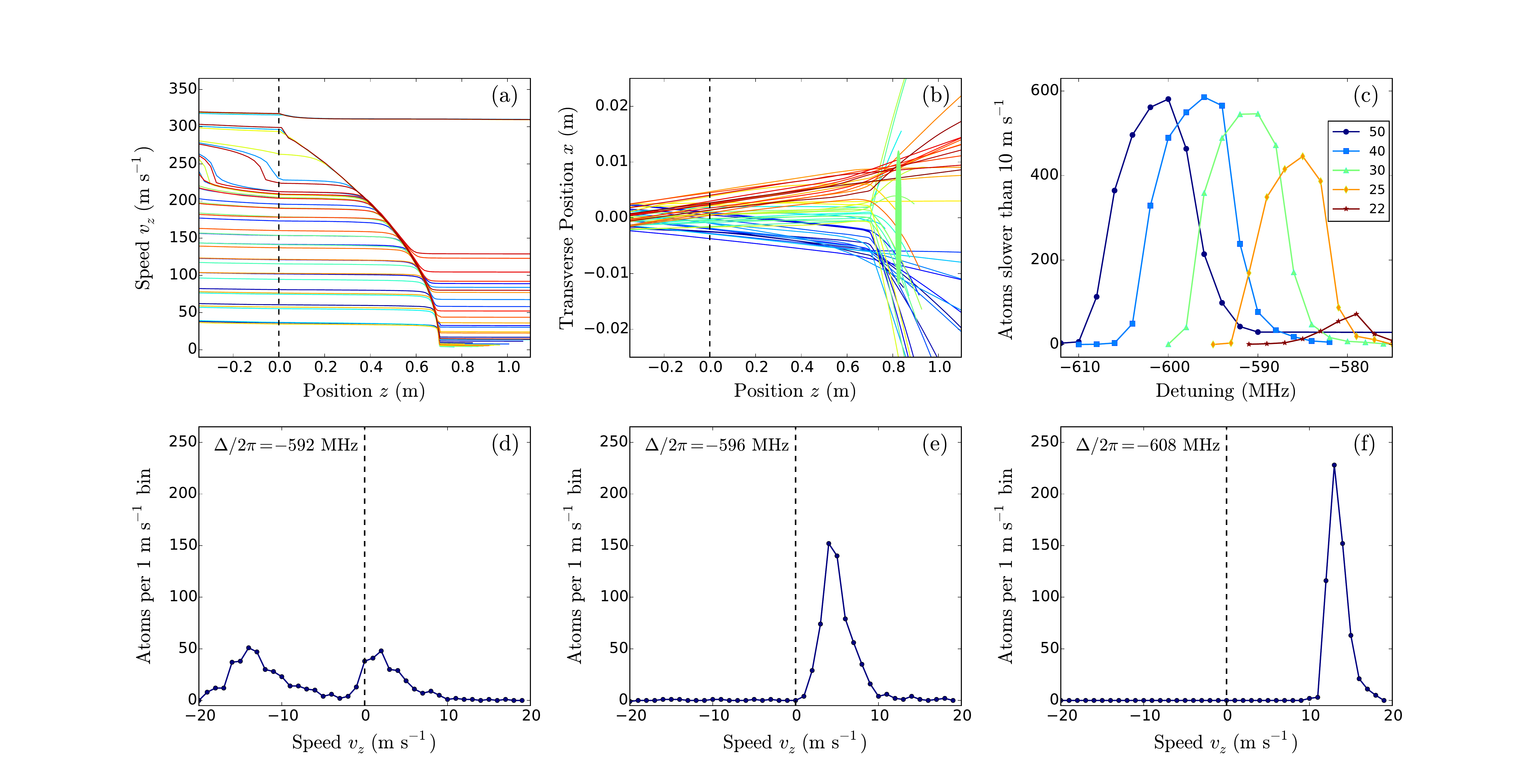}
	\caption{Results of simulations of our slower design after optimisation for Yb. For plots (a) and (b) an initial 50 atoms are launched with an oven effusion half-angle of 5\,mrad. For plots (c) - (f) an initial 1000 atoms are launched directly along the $z$-axis with a Maxwellian speed distribution at 490\,$^{\circ}$C. (a) Longitudinal speeds demonstrating production of a final velocity group centred at 5 m\,s$^{-1}$ and which contains about half of the atoms that  entered with speeds below 300\,m\,s$^{-1}$.  The remaining atoms escape early when their transverse displacements become too large. (b) The transverse displacements for the same 50 atoms; note the difference in the scale of the two axes. The green ellipse indicates the expected MOT capture region, 12.5\,cm from the slower exit at 0.7\,m. This plot highlights the need to minimise the distance from the slower exit to the MOT as the angular effect of the cumulative transverse speeds becomes pronounced when the atom is fully slowed.  (c) Number of atoms exiting with speeds below 10\,m\,s$^{-1}$ as a function of Zeeman laser detuning.  The different curves are simulations with different laser powers: Black = 50\,mW, Blue = 40\,mW, Green = 30\,mW, Orange = 25\,mW, Brown = 22\,mW.  (d) - (f) Final velocity distributions with a laser power of 40\,mW for three laser detunings: (d) -592\,MHz, (e) -596\,MHz, (f) -608\,MHz.}
\label{fig:SimulatorResults}
\end{figure*}

Firstly it became clear that, if extra laser power is available on the slowing transition, it is far more effective to use it in the main slowing beam than to attempt transverse cooling, which is very demanding of laser power \cite{ZeemanTransverse}. However, with a collimated Zeeman laser, increasing the intensity prevents controlled release near the slower exit and pushes the atoms back towards the source as seen in Fig.\,\ref{fig:SimulatorBadReal}. The solution is to focus the beam towards the atomic source so that the intensity $s'(\textbf{r})$ exceeds $s$ at all points along the slower except near the slower exit where one aims for $s'(L )\approx s$. Thus the slower operates with good headroom $\epsilon > 0$ against escape everywhere except at the end where release is desired, and small adjustments to the laser power and/or detuning and/or exit field can be used to tailor the exit speed. In optimising the radial region of good headroom, one must also trade off intensity with beam diameter, as a larger beam improves the uniformity of intensity near the centre.  Table \ref{tab:LaserParameterSummary} shows the optimised laser beam parameters obtained with our simulation, constrained by the maximum available aperture of 16\,mm at the slower exit.

\begin{table}[ht]
	\centering
		\begin{tabular}{ccccc}\hline\hline
			Quantity 	     						& Symbol												& Units					& Cs				& Yb 					\\ \hline
			Laser detuning						& $\Delta$											& MHz						&	-59				& -609					\\ 
			Laser power								& $P$														& mW						& 9.5				&	44							\\ 
			Laser waist								& $w_0$									& $\si{\micro\meter}$		&	32				&	15						\\
			Laser waist position			& -															& m							& 1.2				&	1.2							\\
			Laser radius at slower exit & $w(L)$                 	& mm						& 16				&	16					\\  \hline
		\end{tabular}
	\caption{Optimised design parameters for the Zeeman laser beams with 16\,mm maximum aperture determined by numerical simulation. The laser waist position is measured from the start of the slower towards the oven. See Fig.\,\ref{fig:apparatus}.}  
	\label{tab:LaserParameterSummary}
\end{table}

As well as increasing headroom, the focussing of the Zeeman laser acts to reduce the atomic beam divergence, particularly for slowers of smaller lengths. Ignoring the diffusion due to spontaneous contribution, a laser divergence equal to the oven angle would reduce the oven transverse speed to zero at the slower exit in which case the transverse displacements are roughly halved; this can keep the atoms several mm nearer to the centre axis of the Zeeman laser. Even so, atoms effusing from the oven at too large an angle are lost. In our simulation these losses were negligible for atoms leaving the oven within $\pm 5$\,mrad but then increased rapidly with no atoms being captured for half-angles greater than 10\,mrad.  

To arrive at a suitable magnetic field profile, we calculated the field profile for a defined set of coil windings, tested the profile with the simulation, then fed back the results to refine the windings until we could produce accurately both the large Yb field profile and the smaller Cs profile. We chose to optimise the design manually, so that we could take into account the practical issues such as fitting around the vacuum flanges and fabricating the windings. We did not depart significantly from the functional shape of the basic model of Eq. (\ref{eq:Bprofile}) although there are alternative approaches \cite{Ovchinnikov2007,Ohayon2013}. We paid particular attention to producing a set of windings that could be used to tune exit speed of the atoms whilst at the same time maintaining a near zero field in the nearby MOT region. 

We also studied the acceptable level of ripple in the field profiles where the headroom provided by a focussed laser provides some protection against imperfections in the field gradient. For Yb, deviations needed to be limited to $\pm5$\,G, but for Cs the limit was more exacting at $\pm2$\,G. We note that the headroom $\epsilon$  defined in Eq. (\ref{eq:epsilonApprox}) for fixed saturation parameters $s$ and $s'$ is smaller for Cs : $\Gamma_{\text{Cs}}/{2k_{\text{Cs}}}$ = 2.13\,m\,s$^{-1}$ as opposed to  $\Gamma_{\text{Yb}}/{2k_{\text{Yb}}}$ = 5.78\,m\,s$^{-1}$.

All considerations so far for Cs apply to Zeeman slowing using one of the stretched two-level transitions $(F,m_F)=(4,\pm4)$ to $(F',m'_F)=(5,\pm5)$, but in reality atoms effuse from an oven evenly distributed over the 16  magnetic substates of the $F=3$ and $F=4$ ground states. The fate of the other fifteen states is not immediately obvious because they have different Zeeman shifts and different transition matrix elements for both slowing and repumping. Using the simulation we established that, in the zero-field region before the slower entrance, all atoms initially in $F=3$ can be reliably pumped to $F=4$ and then transferred along the Zeeman manifold to the desired end state i.e. $(F,m_F)=(4,-4)$ for atoms with negative $\mu_{\text{eff}}$ and vice versa. 

\begin{figure*}
	\centering
		\includegraphics[width=1\linewidth]{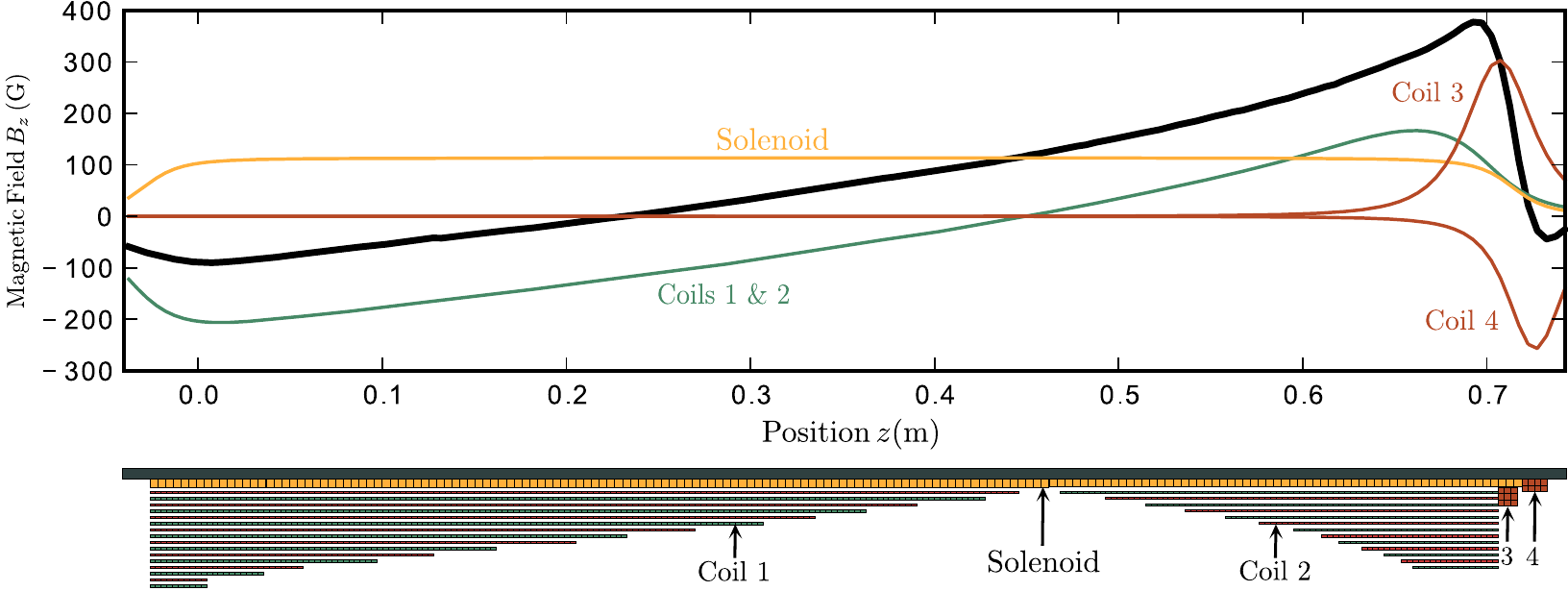}
	\caption{Details of the Zeeman slower windings. The direction of the atomic beam is from left to right. At the top we show the axial magnetic fields of the five separate coils that sum to give the total field profile for Yb (black) which crosses zero near 0.23\,m and at the bottom we show the winding pattern. Green: coils 1 and 2 which produce the main shape are wound with solid 3$\times$1 mm wire. Brown: coils 3 and 4 which control the large end field and its truncation are wound with 3.5\,mm square, Kapton-insulated, hollow water-cooled wire. Orange: the solenoid is wound directly onto a steel former tube (grey) with a single layer of 4.2\,mm square, water-cooled wire. The number of turns can be seen directly in the winding pattern for the solenoid and coils 3 and 4. Coil 1 has 16 layers varying from 153 turns on the innermost layer decreasing to 10 turns on the outermost and coil 2 has 13 layers with 77 turns on the inner layer and 15 on the outer.} 
	\label{fig:windings}
\end{figure*}

Then, once an atom has entered the slower in the desired magnetic state, the losses are quite weak, as they require off-resonant coupling of unintended $\pi$ or $\sigma^{+}$ components in the Zeeman light (perhaps from stray reflections). Using the simulation, we  ascertained that even though such transitions might occur, the atoms are returned to the desired state sufficiently rapidly by the combined action of the repumper beam and the slowing beam. Here `sufficiently rapidly' means that, although the atoms spend some time in the wrong states, there is sufficent headroom $\epsilon$ for the atoms to remain inside the capture envelope. 

After optimising the slower for Yb, the simulation gives performance as illustrated in Fig.\,\ref{fig:SimulatorResults}. Broadly similar behaviour was obtained for Cs with the appropriate Cs parameters.  The main result is that, for an oven effusion half-angle of 5\,mrad, around half of the atoms entering with speeds below the design speeds of 200(300)\,m\,s$^{-1}$ for Cs(Yb)  are entrained into a tight velocity group of FWHM 5\,m\,s$^{-1}$ with a centre speed tunable from 0 to 30\,m\,s$^{-1}$. The tunability can be achieved in three different ways: by varying the laser detuning as shown in Fig.\,\ref{fig:SimulatorResults}, but also by varying the laser power by $\pm 15 \%$ and/or varying the truncation value of the magnetic field by  $\pm$ 8\,G. This three-fold tunability helps to accommodate the imperfections of the real decelerator. The simulation suggests that a reproducible loading of the MOT requires stabilisation of the laser frequency to within $\pm$\,2\,MHz and of the laser power to within $\pm\,2\,\%$.

We end this section by emphasising that the laser beam profile, the atom beam collimation and the slower-to-MOT distance are all important factors in addition to the magnetic field profile.
 
\section{Oven, coils and laser systems}\label{sec:Fabrication}
This section provides practical details of our oven, coils and laser systems.

\textbf{Oven}: The performance of our simulated Zeeman slower was enhanced by reduction of the oven effusion half-angle to less than 10\,mrad, i.e. by increasing the brightness \cite{ZeemanBrightness} of the atomic beam.  A key technique for producing bright, collimated atomic beams \cite{Ramsey1956,Ross1995,Giordmaine1960,Millen2011} is to use an array of narrow capillary tubes as the oven aperture, where the dimensions of the tubes are matched with the mean free path of the effusing gas. To this end we designed an oven aperture comprising 55 parallel steel capillary tubes, each of 0.58\,mm internal diameter and 20\,mm length with a geometric half-angle of 15\,mrad. Narrower tubes would have given an even smaller spread, but with more tubes needed in the array, so our design was a compromise for easier construction and lower risk of accidental clogging. Running the Yb oven at a temperature in the range 420 -- 470\,$^{\circ}$C to create a pressure around  $10^{-3}$\,mbar, we measured a transverse fluorescence linewidth of 14\,MHz FWHM for the 556\,nm Yb transition, which implies through the Doppler shift a HWHM effusion angle of 13\,mrad. From this we estimate that around one third of the Yb atoms effusing from the oven enter the central 5\,mrad cone where they may be successfully slowed if initially below the capture speed.

\textbf{Coils}: In order to produce and tune the optimised magnetic field profiles for Cs and Yb, shown in Fig.\,\ref{fig:FieldProfiles}, we constructed  a set of five coils wound from three types of copper wire. The winding diagram is shown in Fig.\,\ref{fig:windings} along with plots of the individual fields from each coil as well as their sum when running the currents for Yb.

The main shape of the field is created by coils 1 and 2; these are connected in series and run at the same current but the current direction is reversed between them to create two regions with opposite field. The large end field (375\,G for Yb) and sharp drop-off before the MOT region are produced by the two small, high-current coils 3 and 4 with opposing fields  slightly displaced from each other along the $z$ axis. A single-layer solenoid, wound along the entire length of the slower, allows the whole field profile to be shifted up or down to match the Zeeman laser detuning required for Cs or Yb. The direction of current flow in the solenoid can be reversed by an H-bridge switch. An advantage of the zero-crossing profile is that lower currents can be used to generate the total span ($B_L$) of 475\,G. Further details of the coil fabrication can be found in Ref. \cite{Kemp2015}. 

The coils are driven by a low voltage, high current power supply (Agilent model 6681A) capable of 580\,A and set at a constant 2.7\,V. The currents are controlled by banks of high current MOSFETs placed in series with each coil and are stabilised by a servo system using closed loop Hall sensors e.g. (Honeywell CSNJ481). All coils can be switched on/off in less than 20\,ms. The heat dissipated in the coils requires water-cooling for the larger Yb currents but is under 75\,W per coil for all coils. For coils 3, 4 and the solenoid, water is pumped at a pressure of 4\,bar through the 2.75\,mm circular bore of the hollow wires. As a result some cooling is also provided for the innermost layers of coils 1 and 2 by their thermal contact with the solenoid. 

\textbf{Lasers}: A full description of the laser systems used in our experiment may be found in Ref. \cite{Kemp2015,Guttridge2015}. Some features relevant to this paper are as follows. For Yb, the Zeeman laser beam is derived from a Toptica DL Pro (100\,mW) customised for 399\,nm. The red detuning of about 600\,MHz is obtained by passing  a small fraction of the light through two double-passed AOMs at 200 and 100\,MHz; this light is then locked to a fluorescence signal from a separate Yb atomic beam. 
 The Yb slowing beam and the Cs slowing and repumping beams are expanded to large diameters before being mixed on a dichroic mirror and directed into the vacuum chamber through a viewport with a clear diameter of 38\,mm. At that point, the Yb and Cs beams have $1/e^2$ diameters of 28 and 16\,mm respectively and both are focussed along the slower axis, forming waists about 0.3 metres beyond the start of the slower, close to the oven. The beams are slightly clipped (apodised) by the entry viewport, see Fig.\,\ref{fig:apparatus}. Table \ref{tab:CoilsLasers} gives further details of the coils and lasers as actually used.

\begin{table}[h]
	\centering
		\begin{tabular}{cccc}\hline\hline
			Quantity 	     						& Units		& Cs			& Yb 			\\ \hline
			Coils 1 and 2 current 	 	& A				& 1.25		& 3.95	 	\\
			Coil 3 current						& A				& 28.1		& 149			\\ 
			Coil 4 current 						& A				& 23			& 121			\\ 
			Solenoid current					& A				& -20.5		& 38			\\
			Laser waist		  					& $\si{\micro\meter}$		& 62			& 15			\\ 
			Laser waist position			& m				& 0.21		& 0.39			\\ 
			Laser power 							& mW			& 2.95		& 56			\\ 
			Laser detuning						& MHz			& -49.5			& -585				\\
			Repump power							& mW			& 2.3			& -				\\
			Repump detuning						& MHz			& -40			& -			\\ \hline
		\end{tabular}
	\caption{Coil and laser parameters actually used in our Cs-Yb dual-species slower after empirical optimisation for large MOTs. The coils are identified in Fig.\,\ref{fig:windings}. The laser waist positions are measured from the start of the slower towards the oven in both cases. See Fig.\,\ref{fig:apparatus}. The powers are measured just before entry into the vacuum chamber.}
	\label{tab:CoilsLasers}
\end{table}

\section{ Zeeman slower performance}\label{sec:Results}
\begin{figure}
\includegraphics[width=\columnwidth]{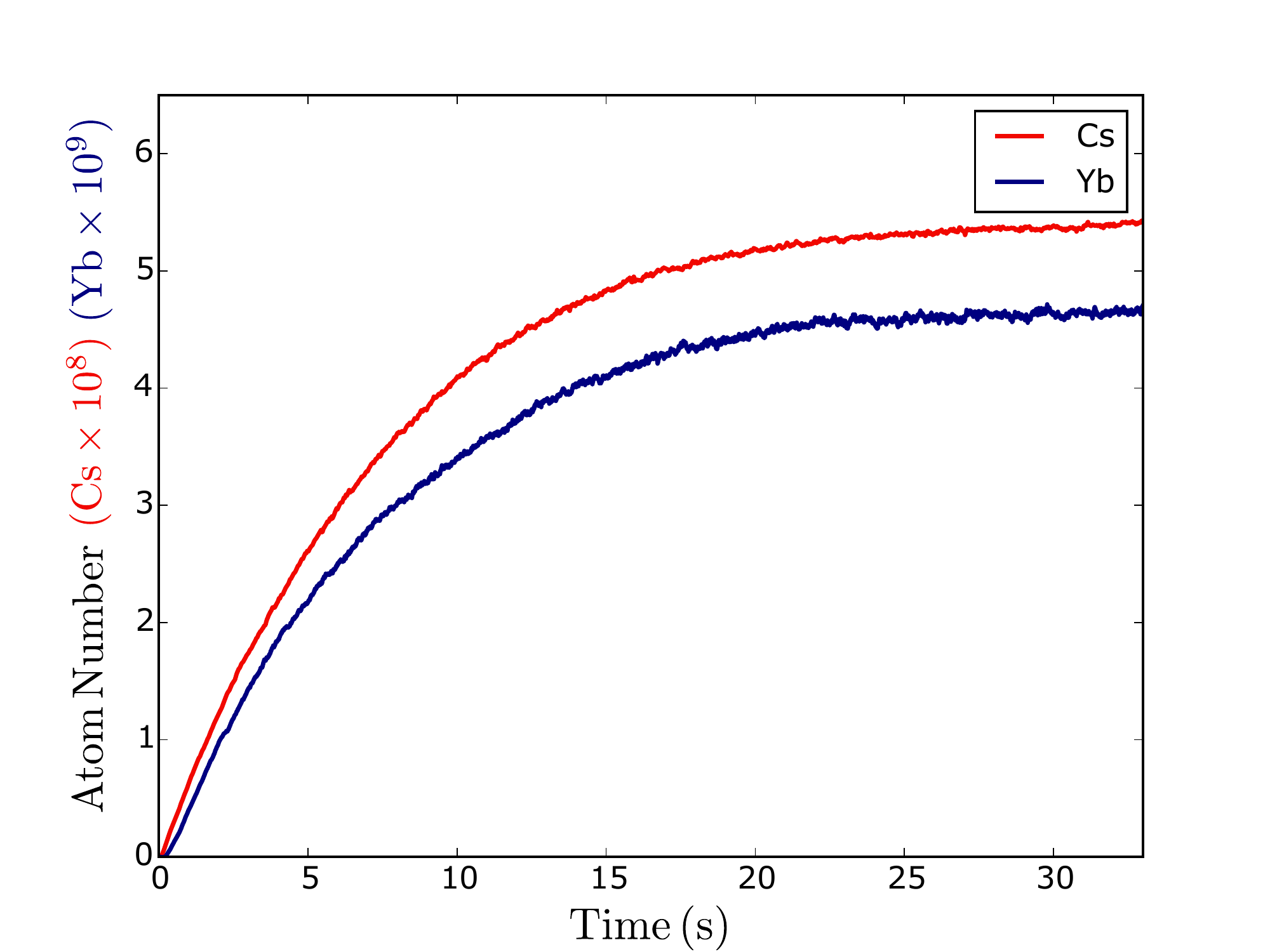}
\caption{MOT loading curves for Cs (red) and Yb (blue). The atom number multiplier is $10^8$ for Cs and $10^9$ for Yb.}
\label{fig:LoadingMOTs}
\end{figure}

As evidence of the success of our design we present in Fig.\,\ref{fig:LoadingMOTs}  MOT loading curves for Yb and Cs, showing loading in a few seconds of more than $10^9$ Yb atoms and more than $10^8$ Cs atoms. The curves are both obtained by monitoring the MOT fluorescence with a lens and photodiode system which has been calibrated using absorption images of the MOT. 

\begin{figure}
\includegraphics[width=\columnwidth]{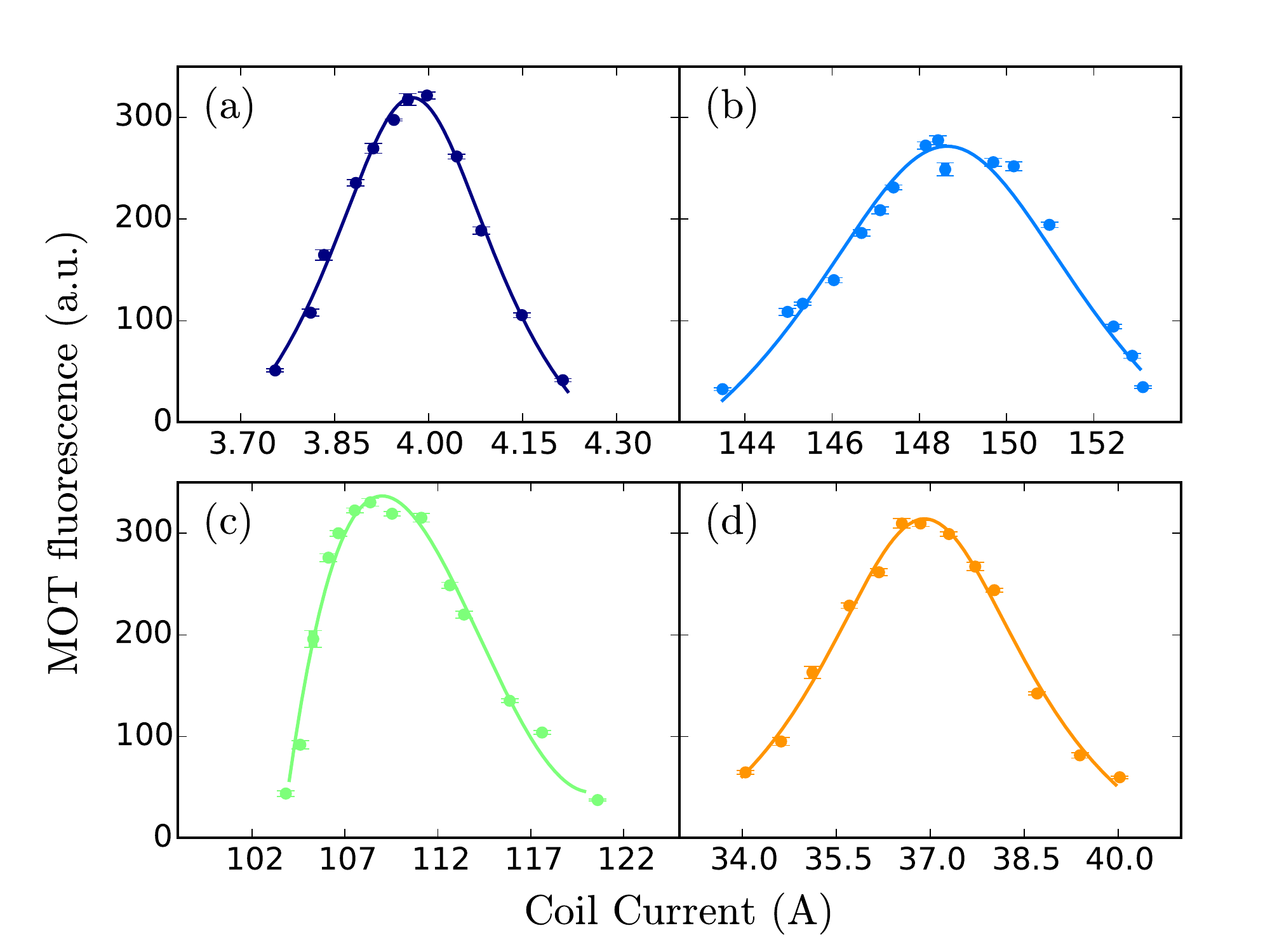}
\caption{The measured Yb MOT fluorescence after three seconds of loading as a function of the Zeeman coil currents, with curves  to guide the eye. Each coil was varied individually whilst the others were maintained near their peak values. a) coils 1 and 2, b) coil 3, c) coil 4 and d) solenoid. The detuning and laser power were -585\,MHz and 65\,mW. }
\label{fig:ZSCoils}
\end{figure}

The Zeeman slower  magnetic field profile was optimised empirically by adjusting the current in each coil and recording the average fluorescence level after loading the MOT for 3 seconds; this is proportional to the MOT loading rate which is in turn proportional to the number of atoms arriving per second with speeds in the MOT capture range. The results, shown in Fig.  \ref{fig:ZSCoils}, demonstrate the sensitivity of the loading of Yb to the currents. Similar results were obtained for Cs at the appropriate smaller currents.

\begin{figure}
\includegraphics[width=\columnwidth]{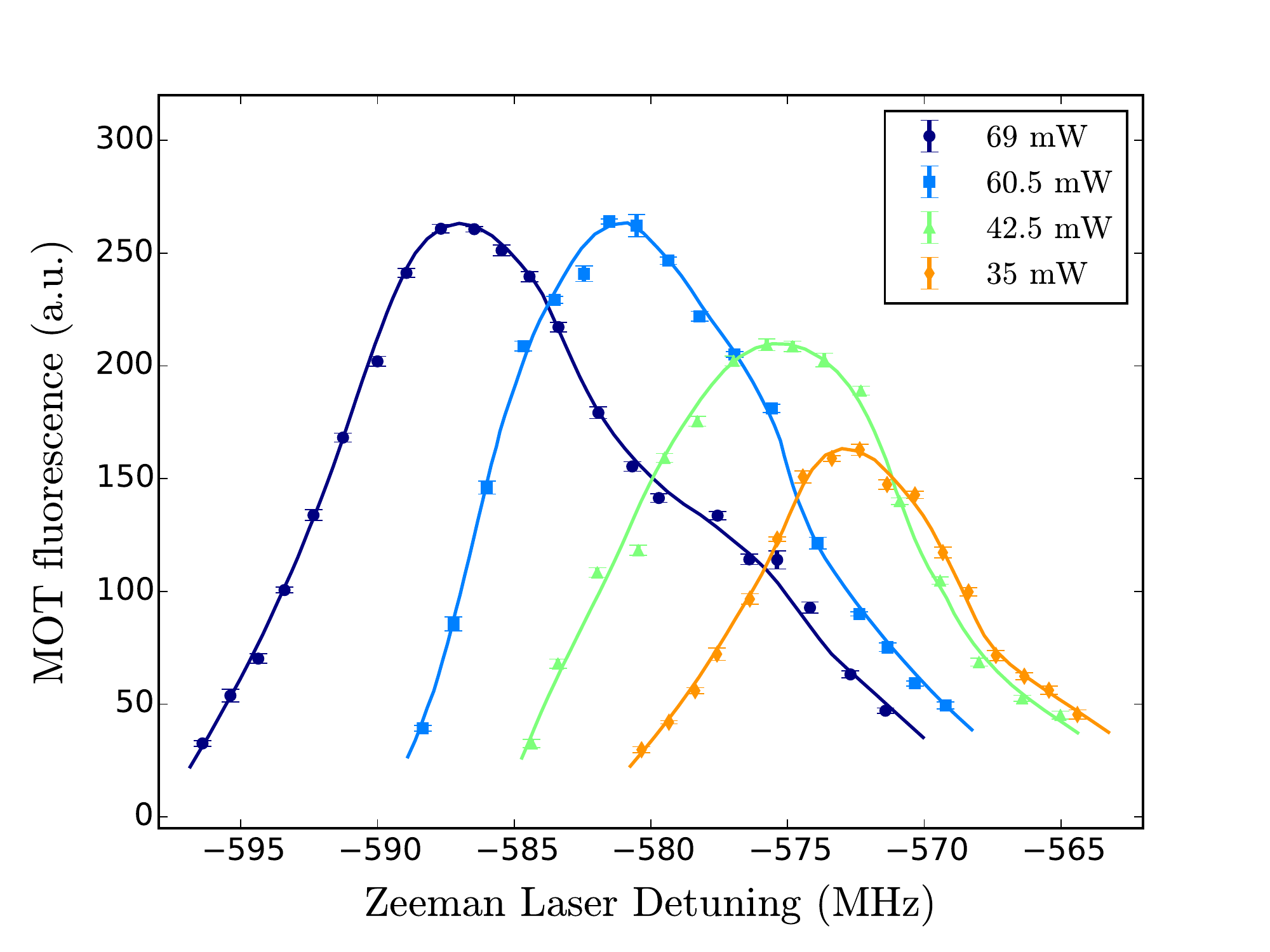}
\caption{The measured Yb  MOT fluorescence after three seconds of loading as a function of the Zeeman laser detuning for four different Zeeman laser powers of 69, 60.5, 42.5 and 35\,mW, with curves added to guide the eye.}
\label{fig:ZSDetuning}
\end{figure}

\begin{figure}
\includegraphics[width=\columnwidth]{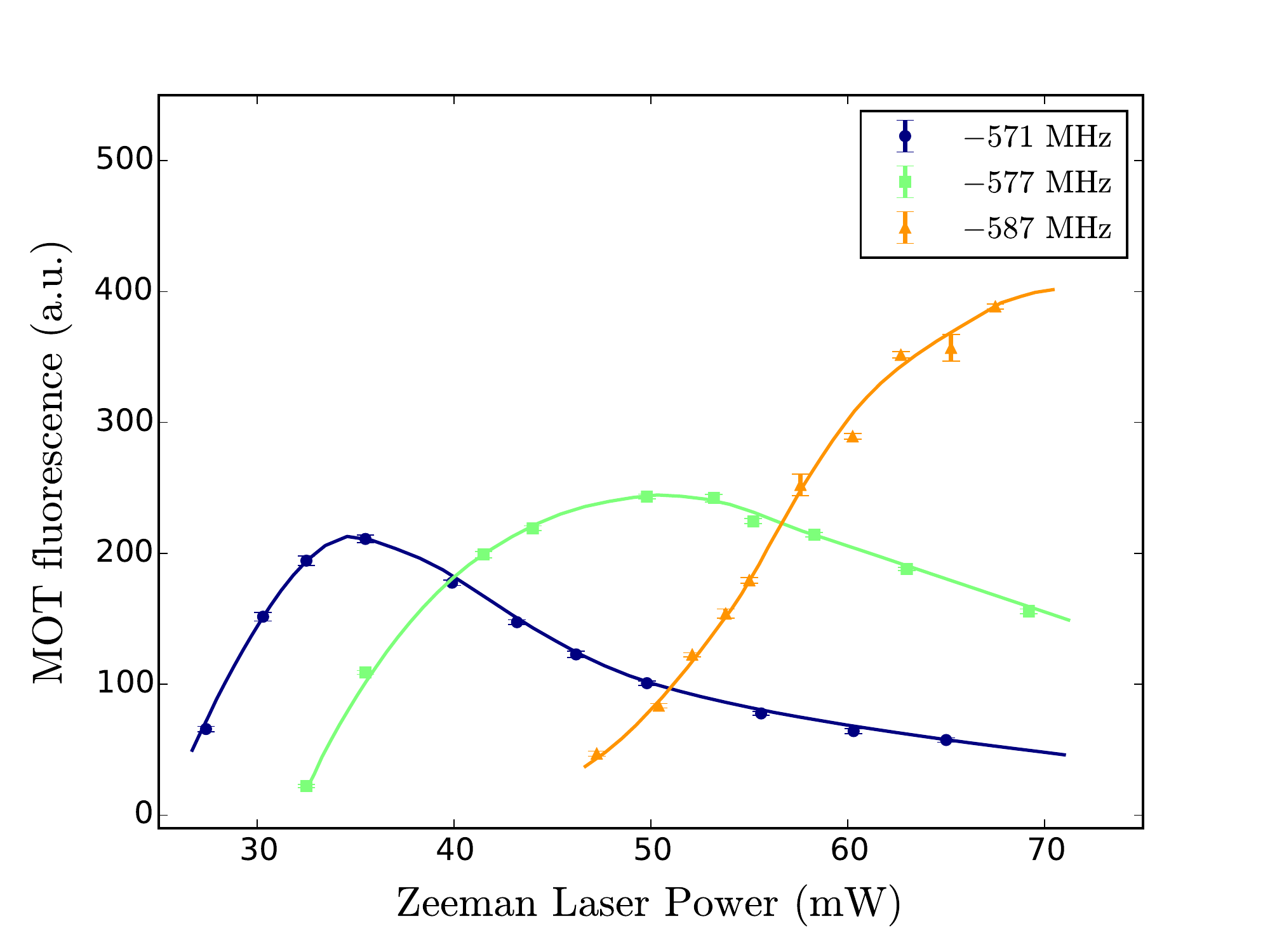}
\caption{The measured  MOT fluorescence after three seconds of loading as a function of the Zeeman laser power for three different Zeeman laser detunings of -571, -577 and -587\,MHz, with curves added to guide the eye.}
\label{fig:ZSPower}
\end{figure}

We also studied the performance of the slower and MOT for both elements as a function of the detuning and power in the Zeeman laser beams.  The results for Yb are shown in Figs. \ref{fig:ZSDetuning} and \ref{fig:ZSPower}. Changing the Zeeman laser detuning changes the exit velocity from the slower and so there is an optimum value for this detuning as shown in Fig.\,\ref{fig:ZSDetuning}. A greater red-detuning increases the exit velocity, and as predicted in Fig.\,\ref{fig:SimulatorResults}(c), optimises the number of atoms at a higher laser power. 

 For a fixed detuning there is an optimum laser power as shown in Fig.\,\ref{fig:ZSPower}. Increasing the power beyond the optimum brings atoms to rest before they reach the MOT. These experimental results agree well with the predictions of our numerical simulation in Fig.\,\ref{fig:SimulatorResults}, apart from an offset of around 10\,MHz\,(Cs) and 20\,MHz\,(Yb) in the optimum laser detuning. Possible explanations for this discrepancy include the back reflection of the Zeeman laser from the end faces of the oven capillary tubes and/or a reduction in intensity of the Zeeman laser due to absorption by the slowed atoms themselves \cite{Ovchinnikov2007}, which increases with atomic density towards the end of the slower. We have not modelled these effects. 

When running the Yb MOT, we could see the 399\,nm Zeeman laser pushing the 556\,nm MOT by 4-5\,mm, even with its detuning of around 20 linewidths. However this effect can be turned off once the MOT is loaded and/or compensated by a set of shim coils normally used to cancel the stray earth field at the MOT. Finally we note that other groups \cite{Kuwamoto1999,Dorscher2013,Hansen2013} have increased their Yb MOT loads by adding sidebands to the 556\,nm MOT light in order to increase the MOT capture speed. In our case, this did not lead to any observable increase in atom number in the MOT as measured by absorption imaging. This may be because our MOT beams, each with intensity of about 40\,$I_{\text{sat}}$ are already sufficiently power-broadened to capture the narrow velocity group delivered by the optimised slower. 

\section{Conclusions}\label{sec:Conclusion}
We have described the design, construction and operation of a versatile dual-species Zeeman slower, optimised for Cs and Yb, but suitable for any of the alkali metals Li, Na, K, Rb, Cs and the alkaline earths Sr and Yb. We reviewed both the simplest and more complete analytic models of Zeeman slowing and also the many practical issues affecting the efficiency of a slower. We also presented the results of a numerical simulation of the slower that elucidates various real-world effects beyond the analytic models and provides solutions to the problems they cause. In particular we have highlighted the usefulness of a large diameter for the Zeeman laser beam and emphasise that the laser beam profile, the atom beam collimation, the slower-to-MOT distance and the magnetic field profile are all important factors.

An optimally designed slower will capture a greater fraction of the flux from an oven with less laser power, resulting in longer oven lifetimes and less contamination of the science chamber by unused flux, as well as less expenditure on lasers. We have demonstrated efficient slowing of both species using only 3\,mW of laser power for Cs and 56 \,mW for Yb. Furthermore in the case of Yb, a MOT based on the narrow intercombination line at 556\,nm has only a small capture speed and benefits enormously from a careful design as we have demonstrated by our experimental results. These results show that our slower, in combination with our MOT beams, can load in a few seconds numbers of atoms in the $10^9$ region for Yb and $10^8$ for Cs. These loading rates are an ideal start for further experiments on ultracold mixtures and molecules.

\section*{Acknowledgements}
This work has been supported by the UK Engineering and Physical Sciences Research Council (grant EP/I012044/1) and the Royal Society. The authors are grateful to Dr Ian Hill at the National Physical Laboratory, Professor Ifan Hughes and Dr Matt Jones at Durham University Physics Department for useful discussions and comments on the manuscript. The data presented in this paper are available from http://dx.doi.org/10.15128/c821gj76b.

\appendix
\section{Theory of Zeeman slowing}\label{sec:Appendix A}
Here we outline, and extend, the theory of Zeeman slowing as originally derived in Refs. \cite{Bagnato1989,Napolitano1990} and further developed in Refs. \cite{Molenaar1997,Ovchinnikov2007}.
The simple model of section \ref{sec:BasicModel} is based on a design deceleration parameter $\eta=s/(1+s)$, where $s$ is the laser saturation parameter $I/I_{\text{sat}}$. This determines the length $L$ and  leads to the velocity trajectory $v(z) = u\sqrt{1-z/L}$ and the associated field profile $B(z) = B_0 + B_L\sqrt{1-z/L}$, given in equations (\ref{eq:vprofile}) and (\ref{eq:Bprofile}). This trajectory is critically unstable for any real-world fluctuations to $v'(z)>v(z)$. The practical resolution to this problem is to increase the laser intensity to a new value $s'>s$ whilst retaining the original field profile $B(z)$ and we now show that this simple empirical adaptation leads to properly stable trajectories.

\begin{figure}
\includegraphics[width=\columnwidth]{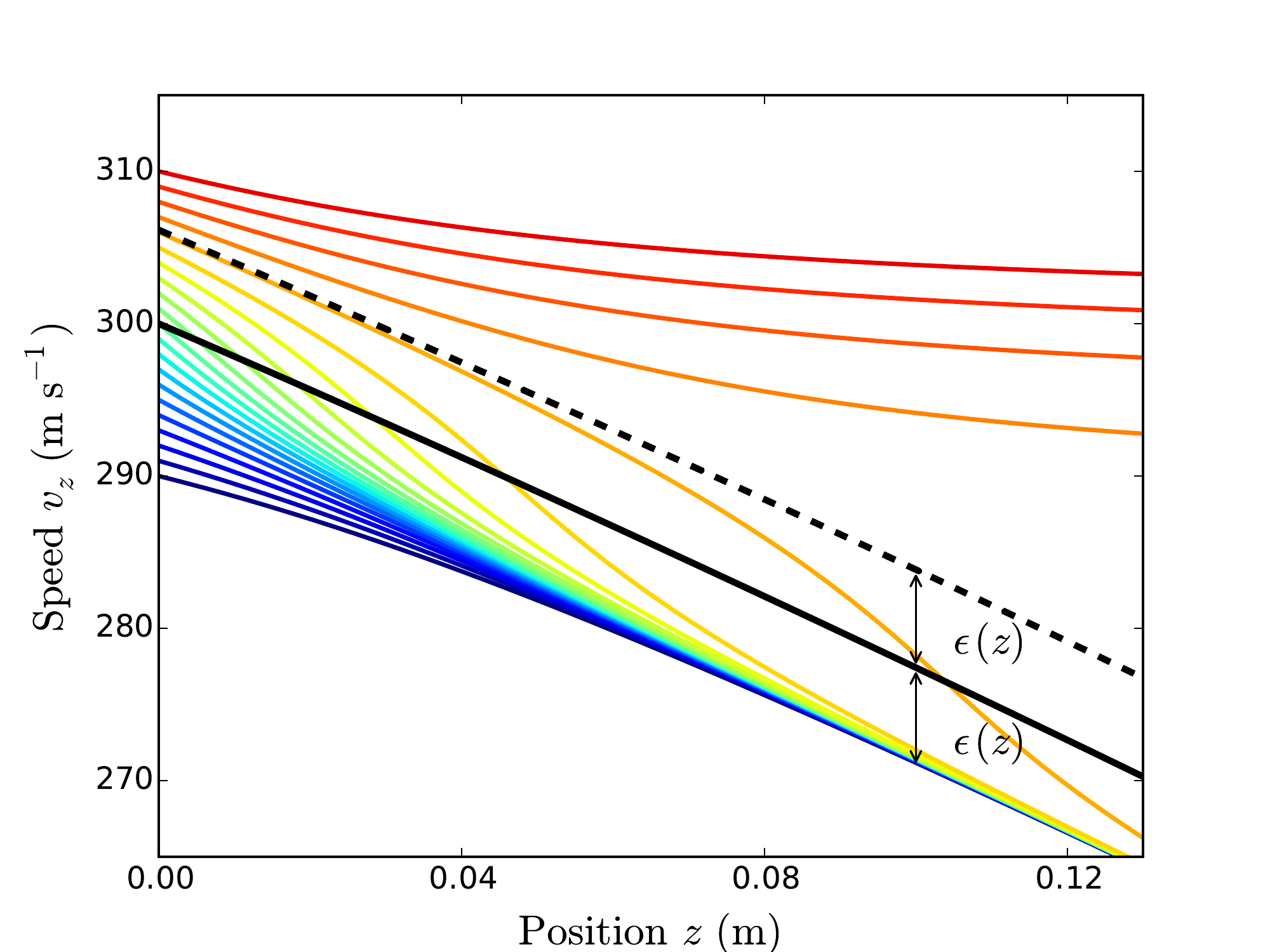}
\caption{Simulation of speed trajectories of Yb atoms entering a slower with design capture velocity $u$ of 300\,m\,s$^{-1}$ and where the intensity is set to produce an offset speed as per Eq. (\ref{eq:epsilonApprox}) of $\epsilon=6.3$\,m\,s$^{-1}$. Twenty-one atoms enter with a uniform spread of speeds in the range 290 - 310\,m\,s$^{-1}$. All captured atoms converge to a trajectory offset by $\epsilon(z)$ below the original design capture envelope, shown by a heavy black line. The convergence band extends upwards to $v(z)+\epsilon$, in this example the dashed black line at 306.3\,m\,s$^{-1}$.}
\label{fig:Stability}
\end{figure}

The higher intensity $s'$ leads to higher deceleration, reducing the velocity below the original design trajectory $v(z)$. The smaller Doppler shift moves the atom to a non-zero detuning $\delta(z)$ and a weaker deceleration, which then allows it to catch up again with the original trajectory. These two opposing effects tend to balance, and the atom `surfs' at a slower velocity $v'(z)$ offset by a small amount $\epsilon(z)$ below $v(z)$, i.e.
\begin{equation}
v'(z) = v(z) - \epsilon(z).  
\label{eq:offset}
\end{equation}
The `surfing condition' is that the new trajectory has a similar slope to the design trajectory i.e. $dv'/dz \approx dv/dz$ or equivalently $d\epsilon/dz \ll dv/dz$ from (\ref{eq:offset}). This picture is confirmed by our numerical simulation and is illustrated in Fig.\,\ref{fig:Stability}. 

To find an analytic expression for $\epsilon(z)$ we first observe that $\delta(z)=-k\epsilon(z)$ on the new trajectory by substituting (\ref{eq:offset}) into Eq. (\ref{eq:LocalDetuning}). We then substitute this and the new intensity $s'$ into Eq. (\ref{eq:ScatteringRate}) to obtain the deceleration:
\begin{equation}
 \frac{dv'}{dt} \equiv v'\frac{dv'}{dz}=-\frac{s'}{1 + s' + 4(k\epsilon(z))^2/\Gamma^2}a_{\text{max}}.
 \label{eq:ScatteringRate2}
\end{equation}
 This deceleration cannot be the same as the constant deceleration of the original design trajectory $v(z,t)$ (precisely because it is offset \cite{Molenaar1997}). Although Equation (\ref{eq:ScatteringRate2}) cannot be integrated analytically, we can eliminate $dv'/dz$ by using the surfing condition above to substitute $dv/dz$ in the place of $dv'/dz$  and $(v'/v \times v)$ for $v'$. This trick introduces the expression $v dv/dz$ on the left-hand side which can then be replaced by $-\eta a_{\text{max}}$ or equivalently $-a_{\text{max}}s/(1+s)$ from Eq. (\ref{eq:Acceleration}). Thus we arrive at a consistency equation for $\epsilon(z)$:
\begin{equation}
 \frac{v'}{v'+\epsilon(z)} \frac{s}{1+s} =\frac{s'}{1 + s' + 4(k\epsilon(z))^2/\Gamma^2},
 \label{eq:Consistency}
\end{equation}
which is a quadratic equation in $\epsilon(z)$ with solutions
\begin{equation}
 \epsilon(z) = \Big[\theta \pm \Big(\theta^2 + \phi^2\Big)^{1/2}\Big],
\label{eq:epsilon}
\end{equation}
where $\theta$ and $\phi$ are speeds given respectively by 
\begin{equation}
 \phi=\frac{\Gamma}{2k} \Big(\frac{s'-s}{s}\Big)^{1/2},
\label{eq:phi}
\end{equation}
\begin{equation}
 \theta=\frac{\Gamma^2}{8k^2}\frac{s'(1+s)}{sv'}.
\label{eq:theta}
\end{equation}

\begin{figure}
\includegraphics[width=\columnwidth]{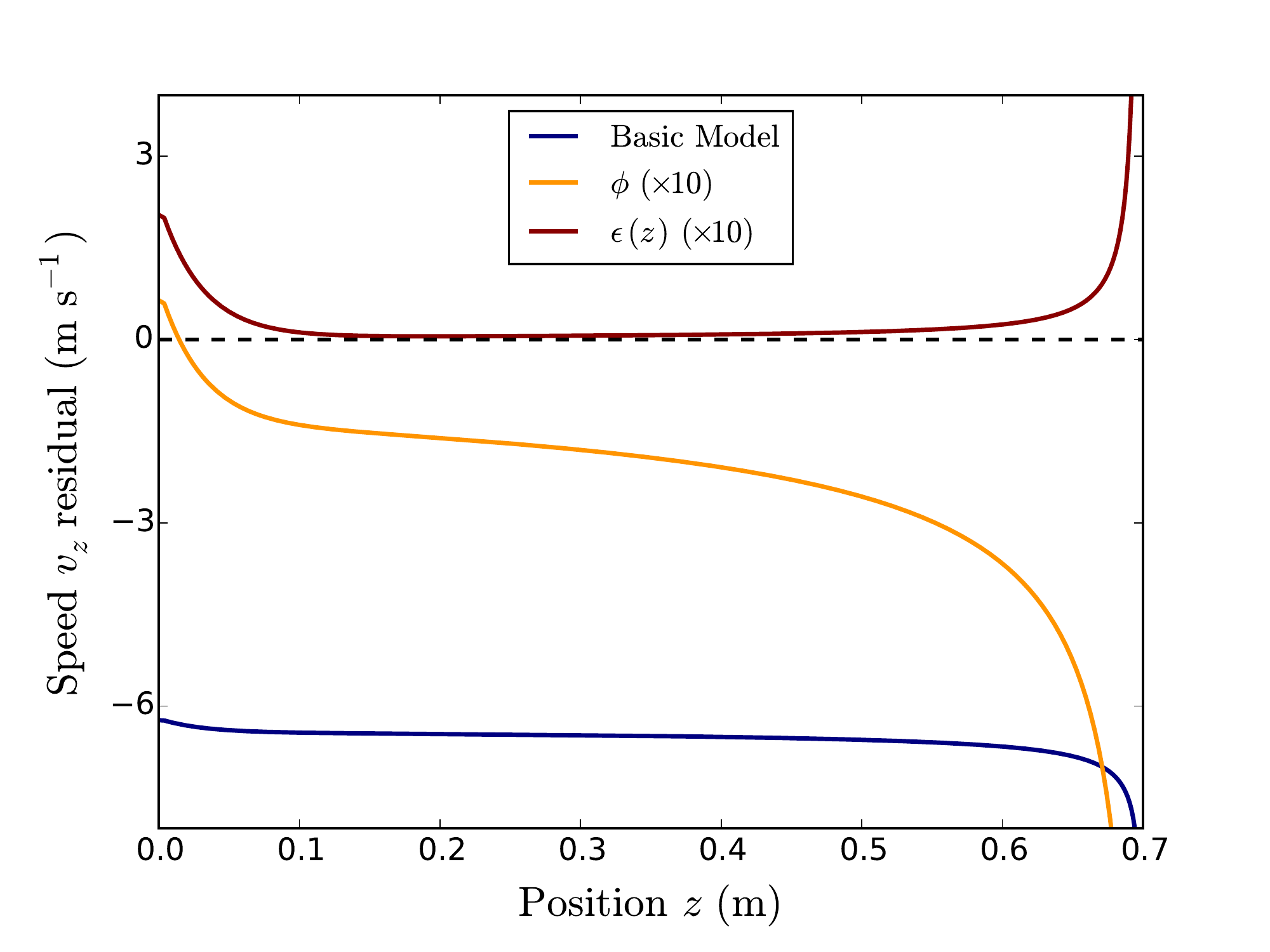}
\caption{Residuals between the simulated trajectory $v_{\text{sim}}(z)$ and trajectories predicted by three models for a Yb atom entering at 300\,m\,s$^{-1}$. Blue: Basic model $v(z)$ as per Eq. (\ref{eq:vprofile}). Orange: the approximate model $v'(z)=v(z)-\phi$ $(\times10)$ of Eq. (\ref{eq:epsilonApprox}). Brown: the improved model $v'(z)=v(z)-\epsilon(z)$  $(\times10)$ of Eq. (\ref{eq:epsilon}). The decelerator conditions are as in Fig.\,\ref{fig:Stability}.}
\label{fig:Appendix}
\end{figure}

When $v'\gg \Gamma/k$,  $\theta$ tends to zero and hence $\epsilon$ approximates to the constant value $\phi$ (see Eq. (\ref{eq:epsilonApprox}) in the main text). That approximation is exactly equivalent to equations [9] and [7] of Refs. \cite{Bagnato1989,Napolitano1990} respectively, derived by considering the dynamics in a uniformly decelerating frame of Ref. \cite{ZeemanDynamics}. The residuals $(\times10)$ between the accurate numerical solution $v_{\text{sim}}(z)$ and the two analytic predictions given  by $\phi$ and $\epsilon$ are shown in Fig.\,\ref{fig:Appendix}, where we see that our new result $\epsilon(z)$ is more accurate over the whole trajectory. Both predictions break down as $v'$ becomes small towards the end of the trajectory, due to the intrinsic assumptions made in each case. We have also confirmed that the residuals are of a similar size for the case when the laser intensity varies along the slower, as with a focussed laser beam. This case has been analysed in detail in Ref. \cite{Ovchinnikov2007}, which points out that for focussed laser beams there can be a more efficient shape for the magnetic profile than Eq. (\ref{eq:Bprofile}). 

Finally we consider the stability of the offset trajectory $v'(z)$. Let there be a small perturbation $\alpha(t)$ away from $v'(t)$ to $v'_{1}(t)$ such that $\alpha(t)=v'_{1}(t)-v'(t)$. Then, by computing $d\alpha/dt$ and employing a little algebra we arrive at
\begin{equation}
 \frac{d\alpha}{dt}=C(\alpha-2\epsilon)\alpha,
\label{eq:alpha2}
\end{equation} 
where $C \approx 4\eta^2 k^2 a_{\text{max}}/(s' \Gamma^2)$ is a positive quantity. For stable trajectories we require negative damping hence we must have $\alpha-2\epsilon<0$ which sets a maximum limit on the allowable fluctuations $\alpha \leq 2\epsilon$. Hence atoms with arrival speeds up to $\epsilon$ greater than the nominal capture speed $u$ can be captured, converging onto a trajectory with a speed $\epsilon$ lower than the design trajectory $v(z)$. This is illustrated by example in Fig.\,\ref{fig:Stability}. Equation (\ref{eq:alpha2}) also permits calculation of the damping times for small fluctuations away from the asymptotic trajectory which, in our case, are a few hundred microseconds. 

To summarise this appendix, the atomic trajectories in a Zeeman slower become stable when the laser intensity is increased over the notional design value, with the atoms then converging onto an offset trajectory $v(z)-\epsilon$, where $\epsilon$ is constant to first order. This phenomenon provides the essential headroom against local technical fluctuations in both the magnetic field profile and laser intensity, and also against Poissonian fluctuations in the local scattering rate. It is this phenomenon that allows Zeeman slowers to work as well as they do.

\bibliography{ZeemanSlowerCsYb}

\begin{thebibliography}{75}
\bibitem [{Zee(note{\natexlab{c}})}]{ZeemanDipolar}%
  \BibitemOpen
  \href@noop {} {} (\bibinfo {year} {Author note}{\natexlab{c}}),\ \bibinfo
  {note} {{T}he radiation pattern of spontaneous emissions for Cs is not
  isotropic, however as this omission from the simulation results in a slight
  overestimate of the transverse spreading, and hence a conservative picture,
  we neglected it.}\BibitemShut {Stop}%
\bibitem [{Zee(note{\natexlab{d}})}]{ZeemanTransverse}%
  \BibitemOpen
  \href@noop {} {} (\bibinfo {year} {Author note}{\natexlab{d}}),\ \bibinfo
  {note} {{T}he transit time of atoms through counter-propagating transverse
  beams filling DN40 viewports is a few hundred microseconds. To reduce an
  initial Yb transverse speed of say 6 ms$^{-1}$ to the Doppler limit of 0.2
  ms$^{-1}$ in this time requires of order 100 mW per 399 nm beam in four
  elliptical-profile beams (40 x 10 mm). It is not possible to cool Cs to the
  Doppler temperature in such a short distance because of its smaller decay
  rate.}\BibitemShut {Stop}%
\bibitem [{Zee(note{\natexlab{e}})}]{ZeemanBrightness}%
  \BibitemOpen
  \href@noop {} {} (\bibinfo {year} {Author note}{\natexlab{e}}),\ \bibinfo
  {note} {{B}y `brightness' we mean atoms per unit area of the source per unit
  solid angle per second. Evaluation of our source brightness is beyond the
  scope of this paper}\BibitemShut {NoStop}%
\end{thebibliography}

 \end{document}